\documentclass[a4,12pt]{article}

\usepackage{amsmath}
\usepackage{amssymb}
\usepackage{theorem}

{
\theoremstyle{plain}
{\theorembodyfont{\rmfamily}
\newtheorem{definition}{Definition}[section]
}
\newtheorem{lemma}[definition]{Lemma}
\newtheorem{proposition}[definition]{Proposition}
\newtheorem{theorem}[definition]{Theorem}
\newtheorem{corollary}[definition]{Corollary}
}
{
\theorembodyfont{\rmfamily}
\newtheorem{remark}[definition]{Remark}
}

\newcommand{\qed}{\hfill $\square$ \\}
\newenvironment{proof}{\noindent \textbf{Proof.}}{\qed \vspace{0.1cm}}

\newcommand{\Ad}{\mathsf{Ad}}
\newcommand{\anal}{\mathrm{A}} 		
\newcommand{\band}{\mathbb{D}}
\newcommand{\bounded}{\mathbb{B}}
\newcommand{\dil}{D} 	

\newcommand{\opend}{\mathsf{End}} 	
\newcommand{\hilbert}{\mathfrak{H}}
\newcommand{\id}{\mathsf{id}}
\newcommand{\Imag}{\mathrm{Im}} 
\newcommand{\ind}{\mathrm{Ind}} 
\newcommand{\ip}[2]{\left( #1 \mid #2 \right)}
\newcommand{\n}{\mathfrak{n}}
\newcommand{\net}{\mathcal{A}}   
\newcommand{\netB}{\mathcal{B}}

\newcommand{\prt}[1]{\left( #1 \right)}
\newcommand{\rot}{R}  	
\newcommand{\trans}{T} 	
\newcommand{\transdil}{\mathcal{P}}   

\newcommand{\neumannB}{\mathcal{B}}
\newcommand{\neumannC}{\mathcal{C}}
\newcommand{\neumannM}{\mathcal{M}}
\newcommand{\neumannN}{\mathcal{N}}
\newcommand{\neumannPsi}{\neumannM(\psi)}

\newcommand{\CC}{\mathbb{C}}

\newcommand{\RR}{\mathbb{R}}

\newcommand{\PSU}{\mathsf{PSU}}

\begin{document}
\title{Endomorphisms on Half-Sided Modular Inclusions}
\author{Rolf Dyre Svegstrup}
\maketitle

\begin{abstract}
In algebraic quantum field theory we consider nets of von Neumann algebras indexed over regions of the space-time. Wiesbrock has shown that strongly additive nets of von Neumann algebras on the circle are in correspondence with  standard half-sided modular inclusions. We show that a finite index endomorphism on a half-sided modular inclusion extends to a finite index endomorphism on the corresponding net of von Neumann algebras on the circle.
Moreover, we present another approach to encoding endomorphisms on nets of Neumann algebras on the circle into half-sided modular inclusions. There is a natural way to associate a weight to a M\"obius covariant endomorphism. The properties of this weight has been studied by Bertozzini, Conti and Longo. In this paper we show the converse, namely how to associate a M\"obius covariant endomorphism to a given weight under certain assumptions, thus obtaining a correspondence between a class of weights on a half-sided modular inclusion and a subclass of the M\"obius covariant endomorphisms on the associated net of von Neumann algebras. This allows us to treat M\"obius covariant endomorphisms in terms of weights on half-sided modular inclusions. As our aim is to provide a framework for treating endomorphisms on nets of von Neumann algebras in terms of the apparently simpler objects of weights on half-sided modular inclusions, we lastly give some basic results for manipulations with such weights.
\end{abstract}

\section{Introduction}

Algebraic quantum field theory is an approach to quantum field theory in which the basic objects of the theory are local algebras of observables associated with bounded regions of space-time. A local algebra associated with a given bounded region of space-time is an algebra of operators corresponding to the physical measurements that can be performed in that region of space-time. Standard introductions to the subject include \cite{Haag-LocalQuantumPhysics}, \cite{Araki} and \cite{Baumgaertel}. 

Commonly we consider states corresponding to particularly simple physical systems in order to obtain a more amenable theory. The Doplicher-Haag-Roberts (DHR) superselection criterion \cite{DoplicherHaagRoberts-Fields} has proven to be fruitful in this respect. In essence it picks out states corresponding to spatially bounded physical systems with no long-range forces. The representations of the net of local algebras whose folia satisfy the DHR superselection criterion are unitarily equivalent to endomorphisms of the local algebras. These endomorphisms are known as DHR endomorphisms and the study of them is central to algebraic quantum field theory.

To model the physical world as we presently understand it, the natural choice of space-time is the four-dimensional Minkowski space. However, theories in lower-dimensional space-times have proven to be both physically and mathematically interesting in their own right. Especially nets of local algebras on the circle have been of growing interest recently. Wiesbrock showed in 1993 that one-dimensional nets of local algebras on the circle satisfying  an assumption known as strong additivity could be completely described by half-sided modular inclusions of factors \cite{Wiesbrock-ConformalQuantumFieldTheory}; a gap in the proof was later filled by Araki and Zsid\'o \cite{ArakiZsido}. 

Wiesbrock's result opens the door to the study of algebraic quantum field theories on the circle in terms of half-sided modular inclusions of factors. In this paper we take the first steps towards developing a theory for M\"obius covariant DHR endomorphisms on half-sided modular inclusions.

We present two main results. First we show an extension theorem for endomorphisms with finite index. Specifically we show that normal, injective endomorphisms with finite index on a half-sided modular inclusion extend to endomorphisms with finite index on the associated net of von Neumann algebras under some light assumptions.
Secondly we present a way of treating M\"obius covariant endomorphisms in terms of weights on half-sided modular inclusions. To any M\"obius covariant endomorphism there is a naturally occurring cocycle satisfying the conditions for being the Connes cocycle derivative of some weight with respect to the vacuum state. Properties of this weight for endomorphisms with finite index has been studied by Bertozzini, Conti and Longo \cite{Longo-KacWakimoto,BertozziniContiLongo-CovariantSectors}. In this paper we construct a method for passing the other way, that is from weights to endomorphisms. This gives us a correspondence between a class of weights on a half-sided modular inclusion and a subclass of the M\"obius covariant endomorphisms on the net of von Neumann algebras associated with the half-sided modular inclusion. Finally, we give the analogues for weights of some basic constructions for endomorphisms; e.g. unitary equivalence, direct sums, and weak conjugates.

Sectionwise the paper breaks down as follows.
In Section \ref{Section: Preliminaries} we recall the basic notions of nets of von Neumann algebras, M\"obius covariant endomorphisms, and half-sided modular inclusions. Next we show in Section \ref{Section: Extending Finite Index Endomorphisms} that any finite index endomorphism on the larger factor of a half-sided modular inclusion extends to a finite index endomorphism on the net of local algebras associated with the half-sided modular inclusion as per Wiesbrock's result.

In \cite{Longo-KacWakimoto} and \cite{BertozziniContiLongo-CovariantSectors}, Bertozzini, Conti and Longo studied for a given M\"obius covariant endomorphism $\rho$, the weight $\psi$ whose Connes cocycle derivative relative to the vacuum state $\omega$ satisfies 
\begin{equation}
\label{Equation: Basic Correspondence}
(D \psi : D \omega)_t = U_\rho(\dil(t)) U(\dil(t))^* 
\end{equation}
where $\dil(t)$ denotes the dilations and $U_\rho$ and $U$ are the usual (projective) representations of the M\"obius group such that $\Ad U_\rho(g) \circ \rho = \rho \circ \Ad U(g)$. In Section \ref{Section: Endomorphisms from Weights} we take a different tack and show conversely that for a suitable weight $\psi$ on the larger factor of a half-sided modular inclusion, we can associate a M\"obius covariant endomorphism $\rho$ on the net of local algebras such that (\ref{Equation: Basic Correspondence}) holds true. We also establish a correspondence between weights and endomorphisms allowing us to treat the latter in terms of weights on half-sided modular inclusions.

Finally in Section \ref{Section: Basic Constructions with Weights} we give the equivalents of basic operations pertaining to endomorphisms in terms of weights including unitary equivalence, direct sums, subrepresentations and weak conjugates. Also we give criteria for finite index and positivity for an endomorphism associated with a weight.


\section{Preliminaries}
\label{Section: Preliminaries}

Below we present the basic objects and results that will form the foundation for later sections. Most notably we introduce nets of von Neumann algebras and their relation with half-sided modular inclusions.

\subsection{Nets of von Neumann Algebras}
We will identify $S^1 := \{z \in \CC \mid |z| =1\}$ with the one-point compactification of $\RR$, $\RR \cup \{\infty\}$, through the stereographic projection,
\begin{equation*}
S^1 \ni z \mapsto x(z) := \frac{1}{i} \frac{z-1}{z+1} \in \RR \cup \{\infty\} .
\end{equation*}
The M\"obius group $\PSU(1,1)$ acts on $S^1$ by
\begin{equation*}
\begin{pmatrix} \alpha & \beta \\ \bar\beta & \bar \alpha \end{pmatrix} \cdot z := \frac{\alpha z + \beta}{\bar\beta z + \bar \alpha} .
\end{equation*}
Two types of elements of note are the translations and dilations.
\begin{equation*}
\begin{array}{lll}
\text{Translations:} \quad & \trans(a) := \begin{pmatrix} 1 + ia/2 & ia /2 \\ -ia /2 & 1 - ia/2 \end{pmatrix}, & a \in \RR
\\
\text{Dilations:} \quad & \dil(t) := \begin{pmatrix} \cosh(\pi t) & \sinh( \pi t) \\ \sinh (\pi t) & \cosh( \pi t) \end{pmatrix}, & t \in \RR
\end{array}
\end{equation*}
By a \emph{proper interval} on $S^1$ we will mean a non-empty, non-dense, open, connected subset of $S^1$. For convenience we will often refer to proper intervals simply as intervals when no confusion can occur.  If $I$ is an interval, we write $I^\bot$ for the interior of its complement in $S^1$. If $I$ and $J$ are intervals such that $I \cap J = \emptyset$, we write $I \bot J$. 

\begin{definition}[Net of von Neumann Algebras]
A \emph{net of von Neumann algebras on $S^1$} is an assignment for each interval $I$ on $S^1$ of a von Neumann algebra $\net(I)$ on a fixed Hilbert space $\hilbert$ satisfying the following axioms:
\begin{itemize}
\item {\rm \bf Isotony:} For intervals $I \subseteq J$ we have $\net(I) \subseteq \net(J)$.
\item {\rm \bf Locality:} If $I$ and $J$ are disjoint intervals, then $\net(I) \subseteq \net(J)'$.
\item {\rm \bf M\"obius Covariance:} There exists a unitary representation $(U,\hilbert)$ of the M\"obius group $\PSU(1,1)$ such that $U(g) \net(I) U(g)^* = \net(g I)$. We write $\alpha_g$ for $\Ad U(g)$.
\item {\rm \bf Positive Energy:} The generator of the one-parameter group $\theta \mapsto U(\rot(\theta))$ is positive.
\item {\rm \bf Existence and Uniqueness of Vacuum Vector:} The subspace of $U$-invariant vectors in $\hilbert$ is one-dimensional. We single out one such vector $\Omega$ of norm one, which will be referred to as the \emph{vacuum vector}. We also require that $\bigcup_I \net(I) \Omega$ is dense in $\hilbert$ where $I$ ranges over the intervals on $S^1$.
\end{itemize}
We will write $\net$ for the family $(\net(I))_I$.
\end{definition}

The above defined concept of a net of von Neumann algebras also appears in the literature under the names of \emph{conformal precosheaf} \cite{GuidoLongo-ConformalSpin} and \emph{local conformal net} \cite{Kawahigashi-Classification} amongst others. The individual von Neumann algebras $\net(I)$ are called \emph{local algebras} and elements of $\bounded(\hilbert)$ belonging to a local algebra are called \emph{local elements}.

We will denote the vector state associated with $\Omega$ by $\omega(\cdot) := \ip{\cdot \, \Omega}{\Omega}$. Usually we will think of this as a vector state on $\net(S_+)$ where $S_+ := \{z \in S^1 \mid \Imag(z) > 0 \}$. For notational ease we will write its restriction to a local algebra $\omega_I := \omega |_{\net(I)}$.

\begin{remark}
If $\net$ is a net of von Neumann algebras, it is often practical to be able to speak of $\net(\mathcal{O})$ for general open sets $\mathcal{O}$. We define $\net(\mathcal{O})$ as follows.
\begin{equation*}
\net(\mathcal{O}) := \bigvee \{ \net(I) \mid \text{$I$ is a proper interval and $I \subseteq \mathcal{O}$} \} .
\end{equation*}
By isotony of the net $\net$, there is no ambiguity when speaking of $\net(I)$ for a proper interval.
\end{remark}

We mention some important consequences of the axioms. All proofs can be found in \cite{Baumgaertel}.

\begin{theorem}[Irreducibility]
If $\net$ is a net of von Neumann algebras on the Hilbert space $\hilbert$ then $\bigvee_I \net(I) = \bounded(\hilbert)$.
\end{theorem}

\begin{theorem}
Every local algebra in a net of von Neumann algebras is a type $III_1$-factor.
\end{theorem}

\begin{theorem}[Haag Duality]
If $\net$ is a net of von Neumann algebras then \emph{Haag duality} holds:
\begin{equation*}
\net(I^\bot) = \net(I)' , \quad \text{$I$ is an interval on $S^1$.}
\end{equation*}
\end{theorem}

\begin{theorem}[Reeh-Schlieder]
The vacuum vector $\Omega$ is cyclic and separating for each local algebra $\net(I)$.
\end{theorem}

We write $r$ for the reflection $S^1 \ni z \mapsto \bar z \in S^1$. For any interval $I$ on $S^1$ we define the \emph{reflection associated with $I$} to be $r_I := g r g^{-1}$ where $g$ is an element of the M\"obius group such that $g S_+ = I$. It is easily checked that $r_I$ is well-defined. We write $\PSU(1,1)_\pm$ for the extended M\"obius group generated by the M\"obius group and the reflections.

Also for any interval $I$ we define the dilations associated with $I$ by $\dil_I(t) := g \dil(t) g^{-1}$ for any $g \in \PSU(1,1)$ such that $g I = S_+$. As with the reflections, it is easy to check that $\dil_I$ is well-defined.

\begin{theorem}[Bisognano-Wichmann]
\label{Theorem: Bisognano-Wichmann}
Let $\net$ be a net of von Neumann algebras. The representation $U$ of the M\"obius group can be extended to the extended M\"obius group in such a way that the net $\net$ is covariant under this representation and
\begin{equation*}
\begin{array}{lll}
U(\dil_I(t)) & = & \Delta_{\net(I)}^{it} \\
U(r_I) & = & J_{\net(I)}
\end{array}
\end{equation*}
where the modular operator and conjugation for $\net(I)$ is with respect to the vacuum state $\omega ( \cdot ) = \ip{ \cdot \, \Omega}{\Omega}$.
\end{theorem}

Finally we mention an important axiom that we will be needing.

\begin{definition}[Strong Additivity]
A net of von Neumann algebras $\net$ is \emph{strongly additive} if for any interval $I$ and point $z \in I$, we have $\net(I) = \net(I_1) \vee \net(I_2)$ where $I_1$ and $I_2$ are the two connected components of $I \setminus \{z\}$.
\end{definition}

To any net of von Neumann algebras $\net$, we can associate the \emph{global $C^*$-algebra $C^*(\net)$} with the following properties \cite{Fredenhagen-SuperselectionSectors, GuidoLongo-RelativisticInvariance}:
\begin{enumerate}
\item There is an embedding of every local algebra $\net(I)$ into $C^*(\net)$, $i_I : \net(I) \rightarrow C^*(\net)$ such that $i_J|_{\net(I)} = i_I$ whenever $I \subseteq J$.
\item If $(\pi_I)$ is a family of representations, $\pi_I : \net(I) \rightarrow \bounded(\hilbert_\pi)$, then there exists a unique representation $(\pi, \hilbert_\pi)$ of $C^*(\net)$ such that $\pi \circ i_I = \pi_I$.
\end{enumerate}


\subsection{Endomorphisms on Nets of von Neumann Algebras}
Let $\net$ be a fixed net of von Neumann algebras on $\hilbert$. A family of $*$-homomorphisms $\rho = (\rho_I)$, $\rho_I : \net(I) \rightarrow \bounded(\hilbert)$ is said to be \emph{consistent} if $\rho_J|_{\net(I)} = \rho_I$ whenever $I \subseteq J$. By the universal property of the global $C^*$-algebra, $\rho$ induces a representation of $C^*(\net)$. 

If there exists an interval $I$ such that $\rho_{I^\bot} = \id_{\net(I^\bot)}$, then $\rho$ is said to be \emph{localized in $I$}. If $\rho$ is consistent and localized in $I$, then $\rho_J$ is an endomorphism of $\net(J)$ whenever $J \supseteq I$ and $\rho$ is an endomorphism of the global algebra $C^*(\net)$. 

A consistent, localized endomorphism is said to be \emph{transportable} if for every interval $J$ there exists a unitary $u$ such that $\Ad u \circ \rho$ is localized in $J$. Such a unitary is called a \emph{transporter}. A consistent, localized and transportable endomorphism is called a \emph{Doplicher-Haag-Roberts endomorphism}, or simply a \emph{DHR endomorphism}. A unitary equivalence class of DHR endomorphisms is called a \emph{sector}.

The requirement of transportability can be strengthened to a requirement of covariance. If $G$ is a subgroup of the M\"obius group, we say that $\rho$ is \emph{covariant with respect to $G$} if there is a projective representation $(U_\rho, \hilbert)$ of $G$ such that 
\begin{equation*}
U_\rho(g) \rho(a) U_\rho(g)^* = \rho \prt{ U(g) a U(g)^* } , \quad g \in G, \; a \in C^*(\net) .
\end{equation*}
If $G$ is either the translation-dilation subgroup of the M\"obius group or the M\"obius group itself, we say that $\rho$ is \emph{translation-dilation covariant}, resp. \emph{M\"obius covariant}. As projective unitary representations of connected semi-simple Lie groups, such as the M\"obius group, lift to true unitary representations of the universal covering groups \cite{BhaskarBagchi}, we will instead of projective representations of the M\"obius group $\PSU(1,1)$ often consider unitary representations of the universal covering group $\overline{\PSU(1,1)}$ of the M\"obius group. As the translation-dilation group is connected and contractible, it is its own universal covering group. Hence, any projective representation of the translation-dilation group stems from a true unitary representation.

A DHR endomorphism $\rho$ localized in some interval $I$ will restrict to an endomorphism $\rho_J$ of $\net(J)$ for any interval $J \supseteq I$. As $\rho_J$ is given by conjugation with a unitary, the inclusion $\rho_J(\net(J)) \subseteq \net(J)$ is a subfactor. We write the (minimal) index of this $\ind(\rho_J)$. If $\rho$ is covariant, it is easy to check that $\ind(\rho_J) = \ind(\rho_{gJ})$ if $J, gJ \supseteq I$ \cite{GuidoLongo-ConformalSpin}. Thus, we will speak merely of the \emph{index of $\rho$} and write this $\ind(\rho)$.

\subsection{Half-Sided Modular Inclusions}
The concept of half-sided modular inclusions and how to construct nets of von Neumann algebras based on these was first introduced by Wiesbrock in 1993 \cite{Wiesbrock-ConformalQuantumFieldTheory}. The article, however, contained a gap which was finally filled by Araki and Zsid\'o in 2004 \cite{ArakiZsido}. Following \cite{GuidoLongoWiesbrock-Extensions} we give some basic definitions and results.

\begin{definition}[Half-Sided Modular Inclusion]
Suppose that $(\neumannN \subseteq \neumannM, \Omega)$ is a triple where $\neumannN \subseteq \neumannM$ is an inclusion of von Neumann algebras on some Hilbert space $\hilbert$ and that $\Omega \in \hilbert$ is cyclic and separating for both $\neumannN$ and $\neumannM$.
\begin{enumerate}
\item The triple $(\neumannN \subseteq \neumannM, \Omega)$ is said to be \emph{standard} if $\Omega$ is cyclic for the relative commutant of $\neumannN$ in $\neumannM$, $\neumannN^c = \neumannN' \cap \neumannM$.
\item The triple $(\neumannN \subseteq \neumannM, \Omega)$ is said to be \emph{$+$ half-sided modular} (resp. \emph{$-$ half-sided modular}) if $\sigma^\neumannM_t(\neumannN) \subseteq \neumannN$ for all $t \leq 0$ (resp. $t \geq 0$) where $\sigma^\neumannM$ denotes the modular automorphism associated with $(\neumannM, \Omega)$.
\end{enumerate}
\end{definition}

For brevity, we will refer to a triple $(\neumannN \subseteq \neumannM, \Omega)$ which is $\pm$ half-sided modular, as a \emph{$\pm$ half-sided modular inclusion}.

\begin{theorem}[\cite{GuidoLongoWiesbrock-Extensions}]
\label{Theorem: Wiesbrock's Theorem}
Suppose that $(\neumannN \subseteq \neumannM, \Omega)$ is a standard $+$ half-sided modular inclusion. Then there exists a unique strongly additive net of von Neumann algebras $\net$ on $S^1$ for which $\net(]0;\infty[) = \neumannM$, $\net(]1;\infty[) = \neumannN$ and $\Omega$ is the vacuum vector.
\end{theorem}


\section{Extending Finite Index Endomorphisms}
\label{Section: Extending Finite Index Endomorphisms}

If $(\neumannN \subseteq \neumannM, \Omega)$ is a standard half-sided modular inclusion and $\net$ is the associated net of von Neumann algebras as per Theorem \ref{Theorem: Wiesbrock's Theorem}, it can be difficult to determine the endomorphisms on $\neumannM$ that extend to DHR endomorphisms on the net $\net$. To extend a given endomorphism on $\neumannM$ to a DHR endomorphism, we would be forced to come up with transporters for all proper intervals on $S^1$. This can be difficult as there need be no particular relation between the transporters associated with different intervals, and the half-sided modular inclusion is only directly linked to the intervals $]0;\infty[$ and $]1;\infty[$.

The situation is easier if we are lucky enough to deal with a translation-dilation covariant endomorphism which is localized in some interval, say $]a;b[$, such that the translation-dilation group alone give all the necessary transporters. A simple way of making sure that the endomorphism has such covariance is by requiring it to have finite index. This even gives M\"obius covariance.

The basic idea will be to use a finite index condition to obtain two one-parameter groups through the Takesaki Theorem. We will then show that these together generate a representation $U_\rho$ of the translation-dilation group such that $\Ad U_\rho(g) \circ \rho = \rho \circ \Ad U(g)$, allowing us to extend the given endomorphism $\rho$ to a translation-dilation covariant endomorphism on the net $\net$. Using the finite index condition once again will finally yield M\"obius covariance of the extended endomorphism.


\subsection{Endomorphisms and $\alpha$-Cocycles}
\label{Subsection: Endomorphisms and Cocycles}
As mentioned above, a DHR endomorphism is completely known by its transporters. As a special case, a translation-dilation covariant endomorphism $\rho$ localized in some interval $I_0$ for which $\infty \in I_0^\bot$, is completely known by the unitary representation $U_\rho$ of the translation-dilation group $\transdil$ satisfying
\begin{equation*}
\Ad U_\rho(g) \circ \rho = \rho \circ \alpha_g , \quad g \in \transdil .
\end{equation*}

Below we make precise the connection between such transporters and translation-dilation covariant endomorphisms. The basic idea and techniques are borrowed from Guido and Longo \cite{GuidoLongo-RelativisticInvariance} where the case of M\"obius covariant endomorphisms are treated. Having only the translation-dilation group and not the full M\"obius group at our disposal engenders some technical problems, as two arbitrarily given intervals are not in general connected by an element of the translation-dilation group.

\begin{remark}[A Comment on the Global $C^*$-Algebra]
\label{Remark: Comment on Global Algebra}
Recall that $\alpha_g = \Ad U(g)$ satisfies $\alpha_g (\net(I)) = \net(g I)$. Thus by the universal property of the global $C^*$-algebra $C^*(\net)$, $\alpha_g$ induces an automorphism of $C^*(\net)$. We will denote this automorphism $\alpha_g$ as well.

As a technical aside we note that when embedding a local element $x$ of $\net$ into the global algebra $C^*(\net)$, we need to specify the interval in which  $x$ is localized. If $I \subseteq J$ and $x$ is localized in $I$, associating $x$ with either $I$ or $J$ leads to the same embedding in $C^*(\net)$ as can be seen directly from the definition of the global $C^*$-algebra.

Thus if $x$ is localized in intervals $I$ and $J$ and there exists a third interval $K \subseteq I \cap J$ in which $x$ is likewise localized, then the embedding of $x$ into the global $C^*$-algebra is the same whether one considers $x$ associated with $I$ or $J$. There are, however, cases where no such interval $K$ can be found and these have to be treated with more care.
\end{remark}

\begin{definition}[$\alpha$-Cocycle]
A family of unitaries $(z_g)_g$ in the universal $C^*$-algebra $C^*(\net)$ indexed by a subset $\mathcal{O}$ of the M\"obius group is said to be an $\alpha$-cocycle on $\mathcal{O}$ with values in $C^*(\net)$ localized in the proper interval $I_0$ if
\begin{enumerate}
\item $z_g$ is localized in $I_0 \cup g I_0$ whenever this is a proper interval.
\item $z_{gh} = z_g \alpha_g(z_h)$ when $g$, $h$, and $gh$ belong to $\mathcal{O}$.
\end{enumerate}
We will refer to the second requirement as \emph{the cocycle condition}.
\end{definition}

Our main interest will be $\alpha$-cocycles indexed either over the translation-dilation group or the full M\"obius group. 

Given a translation-dilation covariant endomorphism, we can define $z_g = U_\rho(g) U(g)^*$ for any $g \in \transdil$. While $(z_g)$ satisfies the cocycle condition the elements will not all have a unique embedding into the universal $C^*$-algebra $C^*(\net)$ as mentioned in Remark \ref{Remark: Comment on Global Algebra}. The next proposition shows how to get around this problem.

\begin{proposition}
\label{Proposition: Local to Global Cocycle}
Let $I_0$ be a subinterval of $S^1$ whose complement contains $\infty$ and let $\mathcal{O} \subseteq \{g \in \transdil \mid I_0 \cup g I_0 \text{ is a proper interval} \}$ be an open set containing the identity.

If $(w_g)_{g \in \mathcal{O}}$ is an $\alpha$-cocycle on $\mathcal{O}$ with values in $C^*(\net)$ then it extends uniquely to an $\alpha$-cocycle on $\transdil$ with values in $C^*(\net)$. 
\end{proposition}
\begin{proof}
Our first step is to choose an open subset $\mathcal{U}$ of $\mathcal{O}$ which contains the identity and satisfies $\mathcal{U}^2 \subseteq \mathcal{O}$. This ensures that whenever we have elements $g$ and $h$ of $\mathcal{U}$, the $\alpha$-cocycle condition makes sense and is satisfied: $w_{gh} = w_g \alpha_g(w_h)$.

The next step is to note that any element of $\transdil$ can be written as a finite product of elements in $\mathcal{U}$. That is, for a given $g \in \transdil$ we can choose a decomposition
\begin{equation*}
g = \prod_{i=1}^n g_i \quad (g_i \in \mathcal{U}) .
\end{equation*}
To extend the cocycle to all of $\transdil$ we will define $w_g$ as
\begin{equation*}
w_g = w_{g_1} \alpha_{g_1}(w_{g_2}) \cdots \alpha_{g_1 \cdots g_{n-1}}(w_{g_n}).
\end{equation*}
To show that this is independent of the choice of $g_i$'s and thus well-defined, it is enough to show that the change $g_i \rightsquigarrow g_i h$, $g_{i+1} \rightsquigarrow h^{-1} g_{i+1}$ for $h \in \mathcal{U}$ leaves the right-hand side of the above equation unchanged. This will follow if we can show that $w_{g_i} \alpha_{g_i}(w_{g_{i+1}}) = w_{g_i h} \alpha_{g_i h}(w_{h^{-1} g_{i+1}})$. We compute
\begin{equation*}
\begin{split}
w_{g_i h} \alpha_{g_i h} (w_{ h^{-1} g_{i+1} }) 
& = w_{g_i} \alpha_{g_i}(w_h) \alpha_{g_i h} (w_{h^{-1}} \alpha_{h^{-1}} (w_{g_{i+1}})) 
\\
& = w_{g_i} \alpha_{g_i}(w_h \alpha_h(w_{h^{-1}}) w_{g_{i+1}}) 
\\
& = w_{g_i} \alpha_{g_i}(w_{h h^{-1}}) \alpha_{g_i}(w_{g_{i+1}})
\\
& = w_{g_i} \alpha_{g_i} (w_{g_{i+1}}) .
\end{split}
\end{equation*}

Hence, $w_g$ is well-defined on all of $\transdil$. The cocycle property follows directly from the definition:
\begin{equation*}
\begin{split}
w_{gh} & =w_{g_1} \alpha_{g_1}(w_{g_2}) \cdots \alpha_{g_1 \cdots g_{n-1}}(w_{g_n}) \alpha_g(w_{h_1}) \alpha_{g h_1}(w_{h_2}) \cdots \alpha_{g h_1 \cdots h_{m-1}} (w_{h_m})
\\
& = w_g \alpha_g (w_h) .
\end{split}
\end{equation*}

The question remains whether $w_g$ will be localized in $I_0 \cup g I_0$ whenever this is a proper interval. Let $g \in \transdil$ be given such that $I_0 \cup g I_0$ is a proper interval and choose another proper interval $J$ extending $I_0 \cup g I_0$ slightly on both sides. It is then possible to choose a decomposition $g = \prod g_i$ such that $I_0 \cup g_1 I_0 \cup \cdots \cup (g_1 \cdots g_n) I_0$ is contained in $J$. Consequently, 
\begin{equation*}
 w_g = w_{g_1} \alpha_{g_1}(w_{g_2}) \cdots \alpha_{g_1 \cdots g_{n-1}}(w_{g_n})
\end{equation*}
belongs to $\net(J)$. By taking the intersection over such intervals $J$ it follows by continuity from the outside that $w_g$ is localized in $I_0 \cup g I_0$.
\end{proof}

The above proposition tells us that whenever we have a translation-dilation covariant endomorphism suitably localized, we obtain a unique $\alpha$-cocycle on $\transdil$ with values in $C^*(\net)$ similarly localized. This situation is similar to the case of M\"obius covariant endomorphisms presented in \cite{GuidoLongo-RelativisticInvariance} where we would get $\alpha$-cocycles on the M\"obius group.

In the latter case there is a bijective correspondence between cocycles and endomorphisms but in the case of $\alpha$-cocycles on the translation-dilation group, which concerns us, we will need to add an extra assumption to counter the problem that the translation-dilation group cannot transform a given interval into any other arbitrary interval. Colloquially put, the problem is that the translation-dilation group `cannot move points through infinity.'

The extra assumption and its use will be elaborated below.

\begin{proposition}
\label{Proposition: Cocycle to Endomorphism}
Let $I_0$ be an interval whose complement contains $\infty$. 
Suppose that $(w_g)$ is an $\alpha$-cocycle on $\transdil$ with values in $C^*(\net)$ localized in $I_0$ and that $w_{\dil(t)}$ is localized in the smallest proper subinterval $J$ of $S^1$ containing $I_0$, $\dil(t) I_0$, and the connected component of $S^1 \setminus (I_0 \cup \dil(t) I_0)$ containing $\infty$ whenever $\bar{I}_0 \cap \dil(t) \bar{I}_0 = \emptyset$.
Then there exists a unique translation-dilation covariant endomorphism $\rho$ on $S^1$ localized in $I_0$ such that
\begin{equation*}
\rho|_{\net(g I_0^\bot)} = \Ad w_g |_{\net(g I_0^\bot)} .
\end{equation*}
\end{proposition}
\begin{proof}
For each proper interval $J$ on $S^1$ containing $\infty$ there is exactly one $g \in \transdil$ such that $g I_0^\bot = J$ and we can therefore define $\rho$ on all such intervals by $\rho_{g I_0^\bot} := \Ad w_g$. Any proper interval not containing $\infty$ is a subinterval of a proper interval which \emph{does} contain $\infty$ so we will define $\rho$ on the former intervals by restriction. To show that this is well-defined, we assume that $J \subseteq g I_0^\bot \cap h I_0^\bot$ and attempt to show that $\Ad w_g |_{\net(J)} = \Ad w_h|_{\net(J)}$. This will be accomplished by showing that $w_g^* w_h \in \net(J^\bot)$. 

We compute
\begin{equation*}
w_g^* w_h = \alpha_g(w_{g^{-1}}) w_h 
= \alpha_g(w_{g^{-1}} \alpha_{g^{-1}}(w_h)) = \alpha_g(w_{g^{-1}h}).
\end{equation*}
As $w_g$ is localized in $I_0 \cup g I_0$ whenever this is a proper interval, it follows that $w_{g^{-1}h}$ is localized in any proper subinterval of $\RR$ containing $I_0 \cup g^{-1}h I_0$. Consequently, $\alpha_g ( w_{g^{-1}h} )$ is localized in any proper subinterval of $\RR$ containing $g I_0 \cup h I_0$. Thus, in the case that $J$ is a subinterval of the connected component of $g I_0^\bot \cap h I_0^\bot$ containing $\infty$, it follows that $\Ad w_g |_{\net(J)} = \Ad w_h |_{\net(J)}$. 

The fact that we don't have the full M\"obius group available but only the translation-dilation group, forces us to consider separately the case where $J$ is a subinterval of a component of $g I_0^\bot \cap h I_0^\bot$ \emph{not} containing $\infty$. In light of the previous arguments, it is enough to show for $t \in \RR$ such that $\bar{I}_0 \cap \dil(t) \bar{I}_0 = \emptyset$ that if $J$ is a subinterval in the connected component of $I_0^\bot \cap \dil(t) I_0^\bot$ not containing $\infty$, then $\Ad w_{\dil(t)} (x) = x$ for any $x \in \net(J)$. This follows directly by the assumptions in the proposition. Thus, we have a well-defined, consistent endomorphism $\rho$.
 
The translation-dilation covariance of $\rho$ easily follows from the definition:
\begin{equation*}
\begin{split}
\Ad (w_g^*) \rho|_{A(h I_0^\bot)} & = \Ad(w_g^* w_h)|_{A(h I_0^\bot)} 
\\
& = \Ad(\alpha_g(w_{g^{-1}}) w_h)|_{A(h I_0^\bot)}
\\
& = \alpha_g \circ \Ad(w_{g^{-1}h}) \circ \alpha_{g^{-1}}|_{A(h I_0^\bot)}
\\
& = \alpha_g \circ \rho|_{A(g^{-1}h I_0^\bot)} \circ \alpha_{g^{-1}} .
\end{split}
\end{equation*}
The covariance of the $\rho_I$'s which are defined by restriction of some $\rho_{g I_0^\bot}$ are also settled by the above calculation.
\end{proof}

\begin{corollary}
\label{Corollary: Endomorphisms-Cocycles}
Suppose that $I_0$ is a proper interval on $S^1$ whose complement contains $\infty$.

Then each translation-dilation covariant endomorphism on $\net$ localized in $I_0$ gives rise to an $\alpha$-cocycle $(w_g)$ localized in $I_0$ and satisfying the assumptions of Proposition \ref{Proposition: Cocycle to Endomorphism} such that
\begin{equation}
\label{Equation: Cocycle-Endomorphism Correspondence}
\rho_{g I_0^\bot} = \Ad w_g , \quad g \in \transdil .
\end{equation}
Conversely, any $\alpha$-cocycle $(w_g)$ localized in $I_0$ and satisfying the assumptions of Proposition \ref{Proposition: Cocycle to Endomorphism} gives rise to an endomorphism $\rho$ on $\net$ localized in $I_0$ and satisfying (\ref{Equation: Cocycle-Endomorphism Correspondence}).
\end{corollary}


\subsection{Extending Finite Index Endomorphisms}
Recall that we have a standing assumption that $(\neumannN \subseteq \neumannM, \Omega)$ is a standard $+$ half-sided modular inclusion and that $\net$ is the net of von Neumann algebras associated with $(\neumannN \subseteq \neumannM, \Omega)$ as per Theorem \ref{Theorem: Wiesbrock's Theorem}.

Our goal in this section is to give sufficient criteria for endomorphisms on $\neumannM$ to extend to DHR endomorphisms on the net $\net$. We will focus on endomorphisms which are translation-dilation covariant. This in particular includes endomorphisms of finite index, as these are automatically M\"obius covariant \cite{GuidoLongo-RelativisticInvariance}. The main element in the approach is to construct an $\alpha$-cocycle on the translation-dilation group and use the results of the previous section to associate a translation-dilation covariant endomorphism to it.

\begin{remark}[Localization]
In general we will not distinguish between $\neumannN \subseteq \neumannM$ and its embedding in $\net$, $\net(]1;\infty[) \subseteq \net(\RR_+)$ and will therefore continue speaking of $\net(I)$ as a sub-von Neumann algebra of $\neumannM$ whenever $I \subseteq \RR_+$. In this spirit we can say an element $x \in \neumannM$ is localized in an interval $I$ if $x \in \net(I)$ under the usual embedding.

Similarly we will call an endomorphism $\rho \in \opend(\neumannM)$ localized in $I \subseteq \RR_+$ if its restriction to $\net(\RR_+) \cap \net(I)'$ is identity. 
\end{remark}

To properly formulate our results completely within the framework of the half-sided modular inclusion we will introduce notation for $\rho$ being localized in some interval $I$ for which $\bar{I} \subseteq S_+$ as follows.

\begin{definition}[Localization Strictly Within $\neumannM$]
An endomorphism $\rho$ on $\neumannM$ is \emph{localized strictly within $\neumannM$} if $\rho$ is localized in an interval whose closure is contained in $S_+$.
\end{definition}

\begin{lemma}
\label{Lemma: Almost Covariant to Covariant}
Suppose that $\rho \in \opend(M)$ is an irreducible endomorphism localized strictly within $\neumannM$, say in $I_0$. Suppose furthermore that $(v_t)$ and $(w_t)$ are unitary one-parameter groups such that 
\begin{enumerate}
\item $\Ad v_s(\rho(x)) = \rho (\alpha_{\dil(s)}(x))$ for $s \in \RR$ and $x \in \neumannM$.
\item $\Ad w_t (\rho(y)) = \rho (\alpha_{\dil_{]1;\infty[}(t)}(y))$ for $t \in \RR$ and $y \in \neumannM \cap \alpha_{\dil_{]1;\infty[}(-t)}(\neumannM)$.
\item $v(s) U(\dil(s))^*$ belongs to $\net(I_0 \cup \dil(s)I_0)$ for $t$ belonging to some neighborhood of zero.
\item $w(t) U(\dil_{]1;\infty[}(t))^*$ belongs to $\net(I_0 \cup \dil_{]1;\infty[}(t) I_0)$ for $s$ belonging to some neighborhood of zero.
\end{enumerate}
Then $\rho$ extends to an irreducible translation-dilation covariant endomorphism on $\net$ localized in $I_0$.
\end{lemma}
\begin{proof}
By standard arguments, $\rho$ is an endomorphism of $\net(J)$ whenever $J \subseteq \RR_+$ is a proper interval containing $I_0$ and as usual irreducibility implies that $\rho(\net(J))' \cap \net(J) = \CC 1$.

The key to proving this lemma is to turn the question of extendability into a question of the existence of a suitable $\alpha$-cocycle. To demonstrate the existence of such an $\alpha$-cocycle we will use the two given one-parameter unitary groups $(v_s)$ and $(w_t)$ to construct a cocycle satisfying the assumptions of Proposition \ref{Proposition: Local to Global Cocycle} and then invoke Proposition \ref{Proposition: Cocycle to Endomorphism} to obtain the desired endomorphism on $\net$.

The first step will be to show that the one-parameter groups $(v_s)$ and $(w_t)$ generate a projective representation of the translation-dilation group $\transdil$ in which $(v_s)$ is identified with the dilations associated with $\RR_+$ and $(w_t)$ with the dilations associated with $]1;\infty[$.

To do this it will be enough to show that $(v_s)$ and $(w_t)$ satisfy the proper commutation relations for small $s$ and $t$. The equivalent of the commutation relation 
\begin{equation}
\label{Equation: Translation-Dilation Relation}
\trans(a) \dil(t) = \dil(t) \trans(\exp(2 \pi t) a)
\end{equation}
takes the shape
\begin{equation}
\label{Equation: Local Translation-Dilation Covariance}
\begin{split}
w(s) v(t-s)  = & v(t) w(-\log(1-\exp(2\pi t) + \exp(2 \pi (t-s)))/ 2\pi) 
\\
& \cdot v(\log(1-\exp(2\pi t) + \exp(2 \pi (t-s)))/ 2\pi)
\end{split}
\end{equation}
for sufficiently small $s$ and $t$. In this particular case `sufficiently small' means $\exp(2 \pi t)(1-\exp(2\pi(s)) \leq 1$ which is satisfied for $(s,t)$ in a neighborhood of $(0,0)$. This requirement ensures that both $a$ and $\exp(2 \pi t)a$ appearing in the former commutation relation are strictly less than one, making it possible to use the equation $T(1 - \exp(2 \pi t)) = \dil(t) \dil_{]1;\infty[}(-t)$.

We set $V(\dil(t)) := v(t)$ and $V(\dil_{]1;\infty[}(s)) := w(s)$. If $g_i \in \transdil$, $i=1, \ldots, n$, is a collection of elements of the form $\dil(t)$ or $\dil_{]1;\infty[}(s)$, where $t$ and $s$ may depend on the index, then we write by abuse of notation $z(g_1 \cdots g_n)$ for the element $V(g_1) \cdots V(g_n) U(g_1 \cdots g_n)^*$. We will show in a moment that $z(g)$ does not depend on the choice of decomposition of $g$. For now we note by the assumptions of the lemma that $z(g_1 \cdots g_n)$ is localized in $I_0 \cup g_1 I_0 \cup g_1 g_2 I_0 \cup \cdots \cup g_1 \cdots g_n I_0$.

Now, choose a neighborhood $\mathcal{O}$ of $(0,0) \in \RR^2$ and a proper interval $J \supseteq I_0$ such that the manipulations in (\ref{Equation: Translation-Dilation Relation}) keep $I_0$ inside $J$ and $J$ inside $\RR_+$ for $(s,t) \in \mathcal{O}$. For convenience, let $z_1$ denote the $z(g_1 g_2)$ corresponding to the left-hand side of (\ref{Equation: Local Translation-Dilation Covariance}) and let $z_2$ denote the $z(g_3 g_4 g_5)$ corresponding to the right-hand side. As the manipulations will map $\net(I_0)$ into $\net(J)$, the element $z_1^* z_2$ is localized in $J$. Therefore to show the validity of (\ref{Equation: Local Translation-Dilation Covariance}) we need only show that $z_1^* z_2$ belongs to $\rho(\net(J))'$.

But by our choice of $J$ and the assumptions of the lemma, it follows that $\Ad(z_1^* z_2) \rho(x) = \rho(x)$ for $x \in \net(J)$.

In conclusion, we get a projective representation $g \mapsto V(g)$ of the translation-dilation group $\transdil$. As $\transdil$ is its own universal covering group we can assume that the representation is a unitary representation. Then defining $z(g) = V(g) U(g)^*$ as before, but this time without abuse of notation, gives us an $\alpha$-cocycle localized in $I_0$ by Proposition \ref{Proposition: Local to Global Cocycle}.

Finally, we want to employ Proposition \ref{Proposition: Cocycle to Endomorphism} to obtain a translation-dilation covariant endomorphism on $S^1$ localized in $I_0$. To do so we must first check that $z(\dil(t))$ is localized in the smallest subinterval $J$ of $S^1$ containing $I_0$, $\dil(t) I_0$, and $\infty$ whenever $\bar{I}_0 \cap \dil(t) \bar{I}_0 = \emptyset$. This, however, is an immediate consequence of the third assumption of this lemma.

Hence, Proposition \ref{Proposition: Cocycle to Endomorphism} provides us with a translation-dilation covariant endomorphism on $\net$ localized in $I_0$. As it obviously coincides with $\rho$ on $I_0$, it is an extension of $\rho$ as desired.
\end{proof}

\begin{remark}
While the former lemma is phrased for unitary one-parameter groups $(v_t)$ and $(w_t)$ corresponding to dilations for the intervals $]0;\infty[$ and $]1;\infty[$, respectively, obviously we can get an analogous result for unitary one-parameter groups corresponding to two intervals $I \subset J$ with one common boundary point. We omit a rephrasing of the lemma.
\end{remark}

The assumptions of the above lemma may not be trivial to check and we will be focusing our attention on a class of endomorphisms which satisfy these assumptions automatically, namely endomorphisms with finite index.

\begin{theorem}
\label{Theorem: Extending Finite Index Endomorphisms}
Suppose that $\rho \in \opend(\neumannM)$ is a finite index endomorphism strictly localized within $\neumannM$. Then $\rho$ extends to a M\"obius covariant endomorphism on $S^1$.
\end{theorem}
\begin{proof}
First off, as $\rho$ is assumed to have finite index, it can be written as direct sum of irreducible endomorphisms similarly localized and we may therefore assume without loss of generality that $\rho$ is in fact irreducible. Moreover, $\rho$ is normal and injective as $\neumannM$ is a type $III$-factor \cite[Theorem V.5.1]{Takesaki1}.

Call the interval in which $\rho$ is localized $I_0$ and choose a proper interval $I \subseteq \RR_+$ containing $I_0$.  As $\rho(\net(I)) \subseteq \net(I)$ has finite index we can find a faithful, normal conditional expectation $E: \net(I) \rightarrow \rho(\net(I))$. We then choose a faithful, normal state $\phi_0$ on $\net(I)$, say the vacuum state, and let $\phi := \phi_0 \circ E$. This is then a faithful, normal state on $\net(I)$ whose modular group $\sigma^\phi$ leaves $\rho(A(I))$ globally invariant by Takesaki's Theorem.

Moreover, the restriction of $\sigma^\phi$ to $\rho(A(I))$ is $\sigma^{\phi_0}$ \cite[Lemma IX.4.21]{Takesaki2}. Thus, on $\net(I)$ we have that \cite[Corollary VIII.1.4]{Takesaki2},
\begin{equation*}
\sigma^{\phi \circ \rho}_t(x) = \rho^{-1} \circ \sigma^{\phi_0}_t \circ \rho(x), \quad x \in \net(I)
\end{equation*}
or equivalently
\begin{equation*}
\rho \circ \sigma^{\phi \circ \rho}_t(x) = \sigma^{\phi_0}_t \circ \rho(x), \quad x \in \net(I) .
\end{equation*}
Let $u_t := (D \phi : D \omega_I)_t$ and $w_t := (D(\phi \circ \rho): D \omega_I)_t$. The usual property of Connes cocycles implies that
\begin{equation*}
\rho \circ \Ad(w_t) \circ \sigma^{\omega_I}_t (x) = \Ad(u_t) \circ \sigma^{\omega_I}_t \circ \rho(x), \quad x \in \net(I).
\end{equation*}
Letting $z_t := \rho(w_t)^* u_t$ and $\alpha_t := \Ad(U(\dil_I(t))$ where $\dil_I$ is the dilation associated with the interval $I$, then the above can be simplified to the expression
\begin{equation*}
\alpha_t \circ \rho \circ \alpha_{-t}(x) = z_t^* \rho(x) z_t .
\end{equation*}
Next, defining $v_t := U(\dil_I(t)) z_{-t}^*$ gives us a unitary satisfying
\begin{equation*}
\Ad v_t \circ \rho(x) = \rho \circ \alpha_t (x), \quad x \in \net(I) .
\end{equation*}
As $\rho$ is irreducible $(v_t)$ is a projective unitary representation of $\RR$. As usual this lifts to a unitary representation of $\RR$ for which we use the same notation. We also note in passing that if $x \in \net(I) \cap \net(I_0 \cup \dil_I(t) I_0)'$ then 
\begin{equation*}
\Ad \left( v_t U(\dil_I(t))^* \right) (x) = \Ad v_t \rho( \alpha_t (x) ) = \rho(x) = x,
\end{equation*}
so $v_t U(\dil_I(t))^*$ belongs to $\net(I) \cap \net(I_0 \cup \dil_I(t) I_0 )'$. In particular, the covariance property $\Ad v_t \circ \rho = \rho \circ \alpha_t$ holds on all of $\alpha_{-t}(\neumannM)$.

Finally, to wrap up the proof, we take the intervals $]0;\infty[$ and $]a;\infty[$ where $a \in \RR_+$ is chosen such that $I_0 \subseteq ]a;\infty[$ and consider the two corresponding unitary one-parameter groups. We have seen above that these satisfy the assumptions of Lemma \ref{Lemma: Almost Covariant to Covariant} and we therefore get an extension of $\rho$ to a translation-dilation covariant endomorphism on $\net$. As $\rho$ has finite index, it is automatically M\"obius covariant.
\end{proof}

The construction of $v_t$ in the proof of Theorem \ref{Theorem: Extending Finite Index Endomorphisms} is borrowed from \cite{GuidoLongo-RelativisticInvariance} where DHR endomorphisms on the Minkowski space with finite statistics and only countably many sectors are shown to be Poincar\'e covariant.

\begin{corollary}
\label{Corollary: Extending Endomorphisms}
If $\rho \in \opend(\neumannM)$ is an endomorphism localized strictly within $\neumannM$ and which is a direct sum of endomorphisms with finite index, then $\rho$ extends to a M\"obius covariant endomorphism on $S^1$ with the same localization.
\end{corollary}


\subsection{Basic Observations for Finite Index Endomorphisms}
Corollary \ref{Corollary: Extending Endomorphisms} gives us a correspondence between finite index endomorphisms on the circle localized in some interval $I$ satisfying $\bar{I} \subseteq S_+$ and injective, normal endomorphisms localized strictly within $\neumannM$. As the correspondence is given simply by restriction of $\rho$ to $\net(S_+)$, it follows that all manipulations of $\rho$ which can be carried out within $S_+$ are preserved by the correspondence.

In particular, unitary equivalence, direct sums, subrepresentations, left inverses, permutators and the like carry over from the net $\net$ to the half-sided modular inclusion $(\neumannN \subseteq \neumannM, \Omega)$ unchanged.

When considering $\alpha$-induction for a quantum field theoretical net of subfactors $\net \subseteq \netB$ \cite{BoeckenhauerEvans-ModularInvariants,LongoRehren-NetsOfSubfactors}, where both nets are strongly additive nets of von Neumann algebras, the net of subfactors can instead be described by a commuting square of subfactors in which the corners are made up by the factors of the half-sided modular inclusions $\net(]1;\infty[) \subseteq \net(]0;\infty[)$ and $\netB(]1;\infty[) \subseteq \netB(]0;\infty[)$.
As such it is in principle possible to perform $\alpha$-induction within the framework of half-sided modular inclusions. However, the fact that the endomorphisms lifted by $\alpha$-induction are not necessarily well-behaved DHR endomorphisms may complicate matters.


\section{Endomorphisms from Weights}
\label{Section: Endomorphisms from Weights}

As in the previous section, our goal is to find a workable equivalent of M\"obius covariant endomorphisms on a strongly additive net of von Neumann algebras $\net$ in the framework of a half-sided modular inclusion $(\neumannN \subseteq \neumannM, \Omega)$ inducing the net of von Neumann algebras as per Theorem \ref{Theorem: Wiesbrock's Theorem}. While we focused on the endomorphism $\rho$ itself in the previous section, in this section we will try a different tack, instead focusing on a subgroup of the transporters of the endomorphism, specifically $t \mapsto U_\rho(\dil(t))$.

Assume that a M\"obius covariant endomorphism $\rho$ on the net $\net$ is given. Assume furthermore that $\rho$ is localized in some interval $I$ satisfying $\bar{I} \subseteq S_+$. By strong additivity of the net $\net$, the endomorphism $\rho$ is completely determined by its restriction to $\net(I)$. By the assumption $\bar I \subseteq S_+$, there exists $t \in \RR_+$ such that $\dil(t) I \bot I$. For such a $t$ the covariance property of $\rho$ implies that $\rho_I = \Ad z_\rho(\dil(t)) |_{\net(I)}$ where as usual $z_\rho(g) := U_\rho(g) U(g)^*$. Hence, knowledge of a subinterval $I$ of $\RR_+$ in which $\rho$ is localized together with knowledge of the one-parameter group $t \mapsto U_\rho(t)$, give complete knowledge of $\rho$.

More natural than considering $t \mapsto U_\rho(\dil(t))$ might be to consider the associated transporters themselves, $t \mapsto z_\rho(\dil(t)) = U_\rho(\dil(t)) U(\dil(t))^*$. It is easy to check that $t \mapsto z_\rho(\dil(t))$ satisfies the condition of being a Connes cocycle derivative of some weight $\psi$ relative to $\omega$, $(D \psi : D \omega)_t = z_\rho(\dil(t))$. The properties of the weight $\psi$, which contains all information about $\rho$, has been studied by Longo, Bertozzini and Conti \cite{Longo-KacWakimoto, BertozziniContiLongo-CovariantSectors} with the main focus on weights associated with endomorphisms with finite index.

In this section we will be taking the opposite approach. That is, starting with a weight $\psi$ we will construct a M\"obius covariant endomorphism $\rho$ such that $(D \psi : D\omega)_t = z_\rho(\dil(t))$. Two of the most fundamental problems in this approach are determining the class of weights that are associated with M\"obius covariant endomorphisms and the level of redundancy there is, that is, when will two weights give rise to the same endomorphism.

As always we fix a standard half-sided modular inclusion $(\neumannN \subseteq \neumannM, \Omega)$ and associate to it a net of von Neumann algebras $\net$ as per Theorem \ref{Theorem: Wiesbrock's Theorem}.


\subsection{Weak M\"obius Covariance}
In Section \ref{Subsection: Endomorphisms and Cocycles} we gave a correspondence between translation-dilation covariant endomorphisms and $\alpha$-cocycles on the translation-dilation group $\transdil$ based on work by Guido and Longo \cite{GuidoLongo-RelativisticInvariance}. Given an endomorphism $\rho$ on $\neumannM$ with finite index, we were then able to associate to it an $\alpha$-cocycle on the translation-dilation group, and the correspondence of $\alpha$-cocycles with endomorphisms then provided us with an extension of $\rho$ to the net $\net$ with finite index.

A key element in the above argument was that we knew \textit{a priori} that $\rho$ had finite index, wherefore it could be written as a direct sum of irreducible endomorphisms. The irreducibility was necessary to get an $\alpha$-cocycle. In this section, however, we will not have finite index \textit{a priori} and will therefore not be able to reduce the problem to the case of irreducible endomorphisms. Thus when faced with a family of unitaries $(v_g)_{g \in \transdil}$ satisfying 
\begin{equation*}
\Ad v_g \circ \rho = \rho \circ \alpha_g,
\end{equation*}
we can no longer conclude that $g \mapsto v_g$ is a projective representation of the translation-dilation group $\transdil$.

The properties that concern us, however, are unaffected by the lack of continuity of $g \mapsto v_g$ as we will see in this and the following section.

\begin{definition}[Weak M\"obius Covariance]
A DHR endomorphism $\rho$ on $S^1$ is said to be \emph{weakly M\"obius covariant} if there exists a family of unitaries $(U_\rho(g))_g$ indexed over the M\"obius group such that $\Ad U_\rho(g) \circ \rho = \rho \circ \alpha_g$.
\end{definition}

Replacing the M\"obius group with the translation-dilation group in the definition above instead gives us what we will call \emph{weak translation-dilation covariance}.

\begin{remark}
If $\rho$ is a weakly M\"obius covariant endomorphism localized in $I_0$ then for any elements $g$ and $h$ of the M\"obius group, $\Ad U_\rho(gh) \circ \rho = \Ad (U_\rho(g) U_\rho(h)) \circ \rho$. Consequently, the element $U_\rho(gh)^* U_\rho(g) U_\rho(h)$ commutes with the image of $\rho$, i.e., 
\begin{equation*}
U_\rho(gh)^* U_\rho(g) U_\rho(h) \in \rho(\net(I_0))' \cap \net(I_0).
\end{equation*}
Thus, $U_\rho$ is a unitary representation of the M\"obius group modulo $\rho(\net(I_0))' \cap \net(I_0)$ in the sense that 
\begin{equation*}
U_\rho(gh) = x_{g,h} U_\rho(g) U_\rho(h) \text{ for some } x_{g,h} \in \rho(\net(I_0))' \cap \net(I_0).
\end{equation*}
For irreducible endomorphisms, the notion of weak M\"obius covariance coincides with that of ordinary M\"obius covariance.
\end{remark}

We found in Section \ref{Subsection: Endomorphisms and Cocycles} that there is a bijective correspondence between $\alpha$-cocycles and M\"obius covariant endomorphisms. In this section we will find a similar correspondence with one part being played by the weakly M\"obius covariant endomorphisms. The other part will be played by $\alpha$-cocycles modulo some von Neumann algebra, to be defined next.

\begin{definition}[$\alpha$-Cocycle Modulo $\neumannB$]
\label{Definition: Weak Alpha-Cocycle}
Let $I_0 \subseteq S^1$ be a proper interval and $\neumannB$ a sub-von Neumann algebra of $\net(I_0)$. An \emph{$\alpha$-cocycle modulo $\neumannB$ localized in $I_0$} is a family of unitaries $(z_g)$ indexed over the M\"obius group satisfying
\begin{enumerate}
\item $z_g \in \net(I_0) \vee \net(g I_0)$ .
\item $z_{gh} = x_{g,h} z_g \alpha_g(z_h)$ for some $x_{g,h} \in \neumannB$ .
\item $\Ad z_g (\net(I_0)) = \neumannB' \cap \net(I_0)$ for some $g$ for which $g I_0 \bot I_0$ .
\end{enumerate}
\end{definition}

\begin{lemma}
\label{Lemma: Trivial Observation}
Suppose that $(z_g)$ is an $\alpha$-cocycle modulo $\neumannB$ localized in $I_0$. Then $\Ad z_g (\net(I_0)) = \neumannB' \cap \net(I_0)$ for every $g$ satisfying $g I_0 \bot I_0$.
\end{lemma}
\begin{proof}
Straightforward.
\end{proof}

The two following propositions will detail how to move back and forth between weakly M\"obius covariant endomorphisms and $\alpha$-cocycles modulo a given von Neumann algebra.

\begin{proposition}
\label{Proposition: Weak Endomorphism to Weak Cocycle}
Suppose that $\rho$ is a weakly M\"obius covariant endomorphism localized in $I_0$. Then $z_g := U_\rho(g) U(g)^*$ defines an $\alpha$-cocycle localized in $I_0$ modulo $\rho(\net(I_0))' \cap \net(I_0)$.
\end{proposition}
\begin{proof}
 Given $x \in \net(I_0)' \cap \net(g I_0)'$. Then
\begin{equation*}
\Ad z_g (x) = \Ad U_\rho(g) (\alpha_{g^{-1}}(x)) = \Ad U_\rho(g) \rho(\alpha_{g^{-1}}(x)) = \rho(x) = x .
\end{equation*}
Thus, $z_g$ belongs to $(\net(I_0)' \cap \net(g I_0)')' = \net(I_0) \vee \net(g I_0)$.

To show that the second requirement of the definition is met, let $g$ and $h$ be elements of the M\"obius group. A quick calculation shows for any local element $x$ that
\begin{equation*}
\begin{split}
\Ad (z_{gh} \alpha_g(z_h)^* z_g^*) \rho(x) &  = \Ad z_{gh} (\alpha_{gh} \circ \rho \circ \alpha_{(gh)^{-1}}(x) ) 
\\
& = \Ad U_\rho(gh) (\rho \circ \alpha_{(gh)^{-1}}(x)) 
\\
& = \rho(x) .
\end{split}
\end{equation*}
Hence, $z_{gh} = x z_g \alpha_g(z_h)$ for some $x \in \rho(\net(I_0))' \cap \net(I_0)$.

Finally we note that for any $g$ such that $g I_0 \bot I_0$, we have $\Ad z_g (\net(I_0)) = \rho(\net(I_0)) = (\rho(\net(I_0))' \cap \net(I_0))' \cap \net(I_0)$.
\end{proof}

\begin{proposition}
\label{Proposition: Weak Cocycle to Weakly Covariant Endomorphism}
Suppose that $(z_g)$ is an $\alpha$-cocycle modulo $\neumannB \subseteq \net(I_0)$ localized in $I_0$. Then there exists a weakly M\"obius covariant $\rho$ localized in $I_0$ giving rise to $(z_g)$ as in Lemma \ref{Proposition: Weak Endomorphism to Weak Cocycle}.
\end{proposition}
\begin{proof}
We aim to define $\rho|_{\net(g I_0^\bot)} = \Ad z_g|_{\net(g I_0^\bot)}$. To show that this gives a well-defined consistent endomorphism on $S^1$ we need to show $z_h^* z_g$ belongs to $\net(g I_0) \vee \net(h I_0)$ for all elements of the M\"obius group $g$ and $h$. 

The cocycle condition implies that $z_h^* = \alpha_h( x z_{h^{-1}})$ for some $x \in \neumannB$. Using that $z_{h^{-1}} \alpha_{h^{-1}}(z_g) = y z_{h^{-1}g}$ for some $y \in \neumannB$ we find that
\begin{equation*}
z_h^* z_g = \alpha_h(x z_{h^{-1}}) z_g =  \alpha_h( x z_{h^{-1}} \alpha_{h^{-1}}(z_g)) =  \alpha_h  ( x y z_{h^{-1}g} )
\end{equation*}
which belongs to $\net(h I_0) \vee \net(g I_0)$. Hence, we obtain a consistent endomorphism $\rho$ on $S^1$.

That $\rho$ is weakly M\"obius covariant can be demonstrated by reapplying the above equation as follows, 
\begin{equation*}
\begin{split}
\Ad z_h^* \circ \rho|_{\net(g I_0^\bot)} & = \Ad (z_h^* z_g)|_{\net(g I_0^\bot)} = \alpha_h \circ \Ad(xy) \circ \Ad z_{h^{-1}g} \circ \alpha_{h^{-1}}|_{\net(g I_0^\bot)} 
\\
& = \alpha_h \circ \rho|_{\net(h^{-1}g I_0^\bot)} \circ \alpha_{h^{-1}} .
\end{split}
\end{equation*} 
Note that Lemma \ref{Lemma: Trivial Observation} is used to eliminate $\Ad(xy)$ in the above computation.
\end{proof}

\begin{remark}[The Case for the Translation-Dilation Group]
\label{Remark: Weak Cocycle to Weakly Translation-Dilation Covariant Endomorphism}
If instead of considering $\alpha$-cocycles indexed over the entire M\"obius group, we consider $\alpha$-cocycles indexed over the translation-dilation group satisfying the requirements of Definition \ref{Definition: Weak Alpha-Cocycle}, then we find a result similar to that of Proposition \ref{Proposition: Weak Cocycle to Weakly Covariant Endomorphism}. The proof can be applied unchanged except that for the translation-dilation covariant case, we have to work with an interval $I_0$ whose complement contains $\infty$.
\end{remark}


\subsection{Weak Conjugates}

In \cite{GuidoLongo-RelativisticInvariance} it is shown how to obtain a weak conjugate of a M\"obius covariant endomorphism based on its associated $\alpha$-cocycle. Following the exposition given in \cite{GuidoLongo-RelativisticInvariance} closely, we will make the necessary generalizations to obtain an  equivalent result for weakly M\"obius covariant endomorphisms.
We start out by recalling some facts about conjugate endomorphisms.

\begin{remark}[Conjugate Endomorphisms]
\label{Remark: Conjugate Endomorphisms}
Assume that $\neumannB \subseteq \neumannC$ is an inclusion of properly infinite von Neumann algebras. Then there exists a normal faithful state $\omega$ on $\neumannC$ represented by a vector in the GNS representation which is cyclic and separating for both $\neumannB$ and $\neumannC$ \cite{Longo-IndexOfSubfactors}. Letting $j_\neumannB$ and $j_\neumannC$ denote the modular conjugations of $\neumannB$ and $\neumannC$ respectively with respect to the aforementioned cyclic and separating vector, we define the \emph{canonical endomorphism} $\gamma : \neumannC \rightarrow \neumannB$ by $\gamma := j_\neumannB j_\neumannC$.

Now suppose that $\rho$ is a unital, injective endomorphism of $\neumannC$ and let $\gamma_\rho$ be the canonical endomorphism associated with $\rho(\neumannC) \subseteq \neumannC$. A conjugate of $\rho$ is then given by $\bar{\rho} = \rho^{-1} \circ \gamma_\rho$.

If $U$ is a unitary that implements $\rho$, that is $\rho = \Ad U$, then $\bar{U} = j_\neumannC (U)$ implements a conjugate $\bar{\rho}$ of $\rho$, $\bar{\rho} = \Ad \bar{U}$. Moreover, every conjugate of $\rho$ is of this form \cite{GuidoLongo-RelativisticInvariance}.
\end{remark}

\begin{definition}[Weak Conjugate Endomorphism]
Suppose that $\rho$ is a consistent endomorphism on the net of von Neumann algebras $\net$ localized in the proper interval $I_0$. A \emph{weak conjugate endomorphism of $\rho$} is a consistent endomorphism $\bar \rho$ localized in $I_0$ such that $\bar{\rho}_I$ is conjugate of $\rho_I$ for every interval $I \supseteq I_0$.
\end{definition}

As usual we extend the M\"obius group and the unitary representation $(U, \hilbert)$ to include reflections associated with proper intervals. In general we will use the notation $r$ or $s$ for such reflections, adding a subscript, $r_I$, if necessary to emphasize the interval with which the reflection is associated. We write $J_I$ for the modular conjugation associated with $(\net(I), \Omega)$ and note that $U(r_I) = J_I$.

\begin{proposition}
Suppose that $(z_g)$ is an $\alpha$-cocycle modulo $\neumannB$ localized in $I_0$. If $h$ is an element of the M\"obius group and $r$ a reflection associated with some interval, then $g \mapsto \bar{z}_g^{h,r} := \alpha_h j_r (z_{rh^{-1}g h r})$ defines an $\alpha$-cocycle modulo $\alpha_h j_r (\neumannB)$ localized in $hr I_0$.
\end{proposition}
\begin{proof}
We fix an element $h$ of the M\"obius group and a reflection $r$ and consider the mapping $g \mapsto \bar{z}^{h,r}_g$.

Firstly, we will deal with the localization. Let $g$ be an arbitrary element of the M\"obius group. Then
\begin{equation*}
\bar{z}_g^{h,r} = \alpha_h j_r (z_{rh^{-1}ghr}) \in \alpha_h j_r \prt{\net(I_0) \vee \net(rh^{-1}ghr I_0) } = \net(hr I_0) \vee \net(g (hr I_0)) .
\end{equation*}
Secondly, as concerns the cocycle condition for $\bar{z}^{h,r}$ we make the following computation.
\begin{equation*}
\begin{split}
\bar{z}_{g_1 g_2}^{h,r} & = \alpha_h j_r \prt{ z_{rh^{-1}g_1 hr r h^{-1}g_2 hr} }
\\
& \in \alpha_h j_r \prt{ \neumannB z_{rh^{-1}g_1 hr} \alpha_{rh^{-1}g_1 hr}(z_{rh^{-1}g_2 hr}) }
\\
& = \alpha_h j_r(\neumannB) \bar{z}_{g_1}^{h,r} \alpha_{g_1} \prt{ \alpha_h j_r (z_{rh^{-1}g_2 hr})}
\\
& = \alpha_h j_r(\neumannB) \bar{z}_{g_1}^{h,r} \alpha_{g_1} \prt{ \bar{z}_{g_2}^{h,r} } .
\end{split}
\end{equation*}
Finally, choose a $g$ such that $g hr I_0 \bot hr I_0$. This clearly implies that $rh^{-1}ghr I_0 \bot I_0$ and we therefore have
\begin{equation*}
\begin{split}
\Ad (\bar{z}_g^{h,r}) \net(hr I_0) & = \Ad \prt{ \alpha_h j_r(z_{rh^{-1}ghr}) } \alpha_h j_r \net(I_0)
\\
& = \alpha_h j_r \prt{ \Ad(z_{r h^{-1}g hr}) \net(I_0) }
\\
& = \alpha_h j_r \prt{ \neumannB' \cap \net(I_0) } 
\end{split}
\end{equation*}
as desired.
\end{proof}

\begin{definition}
For a given weakly M\"obius covariant endomorphism $\rho$ localized in $I$ or equivalently the corresponding $\alpha$-cocycle $(z_g)$ modulo $\rho(\net(I))' \cap \net(I)$, we denote the weakly M\"obius covariant endomorphism associated with $(\bar{z}_g^{h,r})$ by $\bar{\rho}^{h,r}$.
\end{definition}

We fix a weakly M\"obius covariant endomorphism $\rho$ localized in an interval $I_0$ whose complement contains $\infty$.

\begin{lemma}
For any element $h$ of the M\"obius group and reflection $r$, it holds that
\begin{equation*}
\bar{\rho}^{h,r} = \alpha_h \circ j_r \circ \rho \circ j_r \circ \alpha_{h^{-1}} .
\end{equation*}
\end{lemma}
\begin{proof}
\begin{equation*}
\begin{split}
\bar{\rho}^{h,r}_{ghrI_0^\bot} & = \Ad \prt{ \alpha_h j_r (z_{rhg^{-1}hr}) }|_{\net(ghr I_0^\bot)}
\\
& = \alpha_h j_r \circ \Ad(z_{rh^{-1}ghr})|_{\net(rh^{-1}ghr I_0^\bot)} \circ j_r \alpha_{h^{-1}} 
\\
& = \alpha_h j_r \circ \rho_{rh^{-1}ghr I_0^\bot} \circ j_r \alpha_{h^{-1}} .
\end{split}
\end{equation*}
\end{proof}

\begin{proposition}
The endomorphisms $\bar{\rho}^{h,r}$ with $h$ ranging over the M\"obius group and $r$ over the reflections are all in the same sector.
\end{proposition}
\begin{proof}
Let $g$ and $h$ be elements of the M\"obius group and let $r$ and $s$ be reflections. Then
\begin{equation*}
\begin{split}
\bar{\rho}^{h,r} & = \alpha_h j_r \rho j_r \alpha_h^{-1} = \alpha_h j_r j_s \alpha_g^{-1} \bar{\rho}^{g,s} \alpha_g j_s j_r \alpha_h^{-1}
\\
& = \alpha_{hrsg^{-1}} \circ \bar{\rho}^{g,s} \circ \alpha_{(hrsg^{-1})^{-1}}
= \Ad( (\bar{z}_{hrsg^{-1}}^{g,s})^*) \circ \bar{\rho}^{g,s} .
\end{split}
\end{equation*}
\end{proof}

\begin{proposition}
For each proper interval $I \supseteq I_0$ there exists a $\bar{\rho}^{h,r}$ such that $\bar{\rho}_I^{h,r}$ is a conjugate of $\rho_I$.
\end{proposition}
\begin{proof}
Let a proper interval $I \supseteq I_0$ be given. As outlined in Remark \ref{Remark: Conjugate Endomorphisms}, the proof will be done if we can produce a unitary $u$ such that $\rho_I = \Ad u$ and show that $\bar{\rho}^{h,r}_I = \Ad j_I(u)^*$ for some $h$ and $r$. Let $r := r_I$ and choose $h$ such that $hr I_0 = I_0$. It then follows that $\Ad z_{rhr}|_{\net(rhrI_0^\bot)} = \rho_{rhrI_0^\bot} = \rho_{r I_0^\bot}$. Noting that $r I_0^\bot \supseteq I$ we define $u := z_{rhr}$ and consequently obtain $\rho_I = \Ad u |_{\net(I)}$.

By the cocycle condition for $\rho$ there exists an $x \in \rho(\net(I_0))' \cap \net(I_0)$ such that $z_{rhr}^* = \alpha_{rhr}(z_{rh^{-1}r}) x$. Using this we compute
\begin{equation*}
j_r(u^*) = j_r \left( z_{rhr}^* \right) = j_r \left( \alpha_{rhr}(z_{rh^{-1}r} \right) x) 
= \alpha_h j_r \left( z_{rh^{-1}h^{-1}hr} \right) j_r(x)
= \bar{z}_{h^{-1}}^{h,r} j_r(x) .
\end{equation*}
For any $a \in \net(I)$ we have $\Ad (j_r(x)) (a) \in \net(I) \subseteq \net(r I_0^\bot) = \net(h^{-1} I_0^\bot)$. Therefore,
\begin{equation*}
\begin{split}
\Ad j_r(u)^*(a) & = \Ad \left(\bar{z}_{h^{-1}}^{h,r} j_r(x) \right) (a)
\\
& = \Ad\left(\bar{z}_{h^{-1}}^{h,r}\right) \Ad\left(j_r(x)\right)(a)
\\
& = \bar{\rho}^{h,r}\left(\Ad (j_r(x)) a\right)
\\
& = \alpha_h j_r \rho j_r \alpha_h^{-1} \left(\Ad j_r(x) a\right)
\\
& = \Ad\prt{\alpha_h j_r \rho(\alpha_{rh^{-1}r}(x))} \bar{\rho}^{h,r}(a).
\end{split}
\end{equation*}
Now, as $x \in \net(I_0)$ and $rh^{-1}r I_0 \bot r h^{-1} I_0 = I_0$, we have $\rho(\alpha_{rh^{-1}r}(x)) = \alpha_{rh^{-1}r}(x)$. Consequently, 
\begin{equation*}
\Ad \prt{\alpha_h j_r \rho(\alpha_{rh^{-1}r}(x))} \bar{\rho}^{h,r}(a)
= \Ad( j_r(x)) \bar{\rho}^{h,r}(a) = \bar{\rho}^{h,r}(a),
\end{equation*}
as $j_r(x) \in \net(r I_0) \subseteq \net(I)'$ and $\bar{\rho}^{h,r}(a) \in \net(I)$.

In conclusion, $\Ad j_r(u)^*|_{\net(I)} = \bar{\rho}^{h,r}_I$ and  $\bar{\rho}^{h,r}_I$ is therefore a conjugate of $\rho_I$.
\end{proof}

\begin{lemma}
If $h$ is an element of the M\"obius group and $r$ a reflection chosen such that $hr I_0 = I_0$, then for any proper interval $I \supseteq I_0$ the endomorphism $\bar{\rho}^{h,r}$ is unitarily equivalent to $j_I \circ \rho \circ j_I$.
\end{lemma}
\begin{proof}
As in the previous proof we find a unitary $u := z_{rhr}$ such that 
\begin{equation*}
\rho = \begin{cases} \Ad u \text{ on } \net(I) \\ \id \text{ on } \net(I^\bot) \end{cases} .
\end{equation*}
Hence, $j_I \circ \rho \circ j_I$ is given by
\begin{equation*}
j_I \circ \rho \circ j_I = \begin{cases} \id \text{ on } \net(I) \\ \Ad j_I(u) \text{ on } \net(I^\bot) \end{cases} .
\end{equation*}
Conjugating $j_I \circ \rho \circ j_I$ with $j_I(u)^*$ then gives
\begin{equation*}
\Ad\left( j_I(u)^* \right) \circ j_I \circ \rho \circ j_I = \begin{cases} \Ad j_I(u)^* \text{ on } \net(I) \\ \id \text{ on } \net(I^\bot) \end{cases} 
\end{equation*}
which we recognize as $\bar{\rho}^{h,r}$.
\end{proof}

\begin{lemma}
For any interval $I$ and any element $g$ of the M\"obius group, $j_I \circ \rho \circ j_I$ is unitarily equivalent to $j_{gI} \circ \rho \circ j_{gI}$.
\end{lemma}
\begin{proof}
\begin{equation*}
\begin{split}
j_{gI} \circ \rho \circ j_{gI} & = \alpha_g \circ j_I \circ \alpha_g^{-1} \circ \rho \circ \alpha_g \circ j_I \circ \alpha_g^{-1}
\\
& = \alpha_g \circ j_I \circ \Ad z_g^* \circ \rho \circ j_I \circ \alpha_g^{-1}
\\
& \sim j_I \circ \rho \circ j_I \circ \alpha_g^{-1}
\\
& \sim j_I \circ \rho \circ j_I .
\end{split}
\end{equation*}
\end{proof}

We sum up our findings in the following proposition.

\begin{proposition}
\label{Proposition: Form of Weak Conjugates}
Suppose that $\rho$ is a weakly M\"obius covariant endomorphism localized in $I_0$.  Then $\rho$ has a weakly M\"obius covariant weak conjugate and for any interval $I$, $j_I \circ \rho \circ j_I$ is in the sector of that weak conjugate.
\end{proposition}


\subsection{Endomorphisms from Weights}
We have now come to the part where we construct M\"obius covariant endomorphisms from weights. While we ultimately want to formulate the results in terms of half-sided modular inclusions, it is convenient, especially in the proofs, to keep working in the setting of nets of  von Neumann algebras.

The construction falls in two parts. The first part is to construct a weakly M\"obius covariant endomorphism associated with a given weight. This part uses techniques similar to those employed in Section \ref{Section: Extending Finite Index Endomorphisms} where we were able to extend an endomorphism $\rho$ on $\net(S_+)$ to the entire net $\net$ provided that there existed a conditional expectation $E : \net(S_+) \rightarrow \rho(\net(S_+))$ with finite index. We will construct an endomorphism on some local algebra $\net(]a;\infty[)$ along with a conditional expectation onto its image. The fact that this conditional expectation may not have finite index is the reason we are only able to obtain \emph{weak} M\"obius covariance \textit{a priori}.

The second part of the construction consists of showing that the obtained weakly M\"obius covariant endomorphism is a direct sum of endomorphisms with finite index and thus \textit{a fortiori} M\"obius covariant. This part relies on proof techniques developed by Bertozzini, Conti and Longo \cite{BertozziniContiLongo-CovariantSectors} in which the existence of a weak conjugate, proved in the previous section, will play a central role.

\begin{definition}[Localized Weight]
\label{Definition: Alpha-Cocycle Modulo Something}
Let $\psi$ be semi-finite, normal, faithful weight on $\neumannM$. We say that \emph{$\psi$ is localized in the proper interval $I_0 \subseteq \RR_+$} if 
\begin{enumerate}
\item $(D \psi: D \omega)_t \in \net(I_0) \vee \net(\dil(t) I_0)$ for all $t \in \RR$.
\item The restriction of $\psi$ to $\neumannPsi := \bigvee_{t\in \RR} \Ad \prt{(D \psi : D \omega)_t} \prt{\net(\dil(t) I_0)' \cap \neumannM}$ is semi-finite.
\end{enumerate}
\end{definition}
Note that $\neumannPsi \subseteq \net(I_0)$.

\begin{remark}[$\neumannPsi$ Globally Invariant under $\sigma^\psi$]
\label{Remark: PsiM Globally Invariant}
By its very definition $\neumannPsi$ is globally invariant under the modular automorphism group of $\psi$:
\begin{equation*}
\begin{split}
\sigma^\psi_s \prt{ \bigvee_t \Ad(D \psi : D\omega)_t \net(\dil(t) I_0)' \cap \neumannM } & = \sigma^\psi_s \prt{\bigvee_t \sigma^\psi_t \net(I_0)' \cap \neumannM }
\\
& = \bigvee_t \sigma^\psi_t \net(I_0)' \cap \neumannM
\\
& = \bigvee_t \Ad \prt{(D \psi: D \omega)_t} \prt{\net(\dil(t) I_0)' \cap \neumannM }.
\end{split}
\end{equation*}
\end{remark}

The existence of a conditional expectation from $\neumannM$ onto $\neumannPsi$ follows from the remark above. It is, however, difficult to construct an endomorphism on $\neumannM$ with image $\neumannPsi$. Instead we will construct an endomorphism on some $\net(]a;\infty[) \subseteq \neumannM$ by conjugation with $(D \psi : D\omega)_{t_0}$ for some suitable $t_0$. The following lemma asserts that this endomorphism has the proper image, i.e., $\neumannPsi \cap \net(]a;\infty[)$.

\begin{lemma}
Suppose that $\psi$ is a weight localized in the interval $I_0 \subseteq \RR_+$ and that $I \subseteq \RR_+$ is an interval containing the closure of $I_0$. In that case, if $t_0 \in \RR$ is chosen such that $\dil(t_0) I_0 \bot I$, then $\Ad ((D \psi : D \omega)_{t_0}) \net(I) = \neumannPsi \cap \net(I)$.
\end{lemma}
\begin{proof}
We start by showing the inclusion $\Ad ((D\psi: D \omega)_{t_0}) \net(I) \subseteq \neumannPsi \cap \net(I)$.

It is a useful to note that $(D \psi : D \omega)_t^* (D \psi : D \omega)_s$ belongs to $\net(\dil(t) I_0) \vee \net(\dil(s)I_0)$ as seen by the following calculation,
\begin{equation*}
\begin{split}
(D \psi : D \omega)_t^* (D \psi: D\omega)_s & = \prt{ (D \psi : D \omega)_s \sigma^\omega_s( (D \psi: D\omega)_{t-s})}^* (D \psi: D\omega)_s 
\\
& = \sigma^\omega_s( (D \psi : D\omega)_{t-s})^* 
\\
& \in \net(\dil(t) I_0) \vee \net(\dil(s)I_0) .
\end{split}
\end{equation*}
Consequently, the von Neumann algebra $\Ad ((D \psi : D\omega)_{t_0}) \net(I)$ is independent of the choice of $t_0$ as long as $\dil(t_0) I_0 \bot I$. As it is contained in $\net(I) \vee \net(\dil(t_0)I_0)$ we conclude by the independence of $t_0$ that it is in fact contained in the smaller von Neumann algebra $\net(I)$, that is $\Ad ((D \psi : D\omega)_{t_0}) \net(I) \subseteq \net(I)$. Secondly, that $\Ad ((D\psi : D\omega)_{t_0}) \net(I)$ is contained in $\neumannPsi$ is an immediate consequence of $I$ being contained in $\dil(s) I_0^\bot$ for some $s \in \RR$. Hence, we have shown the inclusion $\Ad ((D\psi : D\omega)_{t_0}) \net(I) \subseteq \neumannPsi \cap \net(I)$.

The converse inclusion will follow if we can simply show that 
\begin{equation*}
\prt{\Ad ((D \psi: D\omega)_s) \net(\dil(s) I_0^\bot)} \cap \net(I) \subseteq \Ad (D \psi : D\omega)_{t_0} \net(I) 
\end{equation*}
for all $s \in \RR$.
As the only elements of $\net(\dil(s) I_0^\bot)$ to be mapped into $\net(I)$ by conjugation with $(D \psi : D\omega)_s$ are those localized in $I$, it suffices to show that conjugation with $(D\psi: D\omega)_s$ and $(D \psi: D\omega)_{t_0}$, respectively, are identical on $\net(\dil(s)I_0^\bot \cap I)$. This, however, is a consequence of the aforementioned fact that $(D \psi : D\omega)_t^* (D \psi : D\omega)_s$ belongs to $\net(\dil(t)I_0) \vee \net(\dil(s) I_0)$. Thus, we find that $\neumannPsi \cap \net(I) \subseteq \Ad ((D \psi: D\omega)_{t_0}) \net(I)$.
\end{proof}

As shown as a step in the previous proof,
\begin{equation*}
(D \psi : D \omega)_t^* (D \psi : D\omega)_s \in \net(\dil(t) I_0) \vee \net(\dil(s) I_0) .
\end{equation*}
This has the consequence that not only the image of the endomorphism $\Ad (D \psi : D\omega)_{t_0}|_{\net(I)}$ but also the endomorphism itself, is independent of $t_0$ as long as $\dil(t_0) I_0 \bot I$.

\begin{proposition}
\label{Proposition: From Weight to Weakly Covariant Endomorphism}
Let $I_0$ be an interval such that $\bar I_0 \subseteq S_+$. For any weight $\psi$ on $\neumannM$ localized in $I_0$ there exists a unique weakly M\"obius covariant endomorphism $\rho$ localized in $I_0$ such that $\sigma^\psi_t \circ \rho = \rho \circ \sigma^\omega_t$ on $\neumannM$.
\end{proposition}
Many aspects of the proof will closely parallel arguments presented in Section \ref{Section: Extending Finite Index Endomorphisms}. Instead of giving the same arguments again, we will refer back to the relevant proofs and restrict ourselves to dealing with the differences.
\vspace{\baselineskip}

\begin{proof}
As $\psi$ by assumption is semi-finite on both $\neumannM$ and $\neumannPsi$ and the latter as noted in Remark \ref{Remark: PsiM Globally Invariant} is globally invariant under the modular automorphism group of $\psi$, the Takesaki Theorem supplies us with a faithful conditional expectation $E$ from $\neumannM$ onto $\neumannPsi$.

We choose a proper interval $I$ such that $\bar{I}_0 \subseteq I \subseteq ]\delta;\infty[$ for some $\delta > 0$. We also choose a $t_0$ such that $\dil(t_0) I_0 \bot I$. From the previous lemma it follows that we can define an endomorphism $\rho$ on $\net(I)$ with image $\neumannPsi \cap \net(I)$ by $\rho := \Ad (D \psi : D\omega)_{t_0}$. As $\net(I_0)' \cap \neumannM \subseteq \neumannPsi$, it follows that the restriction of $E$ to $\net(I)$ is a faithful conditional expectation of $\net(I)$ onto $\neumannPsi \cap \net(I)$.

We are now in the situation where we have an endomorphism on $\net(I)$ along with a conditional expectation onto its image. If the endomorphism was irreducible we could apply the proof of Theorem \ref{Theorem: Extending Finite Index Endomorphisms} to extend $\rho$ to a M\"obius covariant endomorphism on $S^1$. That $\rho$ might be reducible will present some problems but the basic techniques used in the previous section can still be used. Our first goal is to construct an $\alpha$-cocycle on the translation-dilation group modulo $\rho(\net(I))' \cap \net(I)$ and use this to extend $\rho$ to all of $S^1$ as per Remark \ref{Remark: Weak Cocycle to Weakly Translation-Dilation Covariant Endomorphism}

As in the proof of Theorem \ref{Theorem: Extending Finite Index Endomorphisms} we can find a one-parameter group $(v_t)$ such that $\Ad v_t \circ \rho = \rho \circ \sigma^{\omega_I}_t$. Taking a proper subinterval $J$ of $I$ with one endpoint in common with $I$ and which contains the closure of $I_0$, we can find another one-parameter unitary group $w_t$ such that $\Ad w_t \circ \rho|_{\net(J)} = \rho|_{\net(J)} \circ \sigma^{\omega_J}_t$. We can assume without loss of generality that the endpoint shared by $I$ and $J$ is $\infty$. It is important to note that $v_t \Delta_{\omega_I}^{-it}$ belongs to $\net(I)$ and that $w_t \Delta_{\omega_J}^{-it}$ belongs to $\net(J)$.

We want to show that the one-parameter unitary groups $(v_t)$ and $(w_t)$ together generate a representation of the translation-dilation group up to left multiplication by elements of $\rho(\net(I))' \cap \net(I)$. As in the proof of Lemma \ref{Lemma: Almost Covariant to Covariant}, it is enough to show that $(v_t)$ and $(w_s)$ satisfy a commutation relation of the shape (\ref{Equation: Local Translation-Dilation Covariance}),
\begin{equation}
\label{Equation: Local Weak Translation-Dilation Covariance}
\begin{split}
w(s) v(t-s)  = & v(t) w(-\log(1-\exp(2\pi t) + \exp(2 \pi (t-s)))/ 2\pi) 
\\
& \cdot v(\log(1-\exp(2\pi t) + \exp(2 \pi (t-s)))/ 2\pi) \, ,
\end{split}
\end{equation}
up to left multiplication by elements of $\rho(\net(I))' \cap \net(I)$ for small $s$ and $t$.

For convenience we will write $U_\rho(\dil_I(t))$ for $v_t$ and $U_\rho(\dil_J(t))$ for $w_t$. Let $g_1, \ldots, g_5$ denote the elements of the group generated by $\dil_I(s)$ and $\dil_J(t)$, $s,t \in \RR$, corresponding to the equation (\ref{Equation: Local Weak Translation-Dilation Covariance}). With argumentation similar to that of the proof of Lemma \ref{Lemma: Almost Covariant to Covariant} we find for a suitably chosen `small' interval $K \supseteq I_0$ that
\begin{equation*}
U_\rho(g_5)^* U_\rho(g_4)^* U_\rho(g_3)^* U_\rho(g_1) U_\rho(g_2) \in \rho(\net(K))' .
\end{equation*}
As in the proof of Lemma \ref{Lemma: Almost Covariant to Covariant}, the interval $K$ is chosen such that the successive application of the concerned elements of the M\"obius group, $g_1, \ldots, g_5$, keep $I_0$ within $K$ and $K$ within $S_+$.
Also with argumentation like in the proof of Lemma \ref{Lemma: Almost Covariant to Covariant}, we find that $U_\rho(g_5)^* U_\rho(g_4)^* U_\rho(g_3)^* U_\rho(g_1) U_\rho(g_2)$ belongs to $\net(K)$. By considering smaller and smaller intervals $K \supseteq I_0$ we can conclude that $U_\rho(g)$ is well-defined up to right multiplication with $\rho(\net(I))' \cap \net(I)$ for $g$ in the group generated by $\dil_I(s)$ and $\dil_J(t)$. It is easy to check that conjugation by $U_\rho(g)$ defines an automorphism on $\rho(\net(I))' \cap \net(I)$ wherefore $U_\rho(g)$ is well-defined up to left multiplication of elements in $\rho(\net(I))' \cap \net(I)$. That is, for given $g_1$ and $g_2$ in the group generated by $\dil_I(s)$ and $\dil_J(t)$, 
\begin{equation*}
U_\rho(g_1) U_\rho(g_2) = x_{g_1, g_2}  U_\rho(g_1 g_2) \; \text{ for some } \; x_{g_1, g_2} \in \rho(\net(I))' \cap \net(I) .
\end{equation*}

We now define $z_g := U_\rho(g) U(g)^*$. The aim is to show that $(z_g)$ defines an $\alpha$-cocycle modulo $\rho(\net(I))' \cap \net(I)$ localized in $I_0$ such that we can apply Remark \ref{Remark: Weak Cocycle to Weakly Translation-Dilation Covariant Endomorphism} to extend $\rho$ to all of $S^1$. The previous arguments show that $(z_g)$ satisfies the cocycle requirement of Definition \ref{Definition: Alpha-Cocycle Modulo Something}. By the very definition of the endomorphism $\rho$ and the weight $\psi$, it is clear that $(z_g)$ also satisfies the third requirement of the definition, namely that $\Ad (z_g) \net(I_0) = \rho(\net(I_0))$ for some $g$ such that $g I_0 \bot I_0$.

For the last requirement, i.e., $z_g \in \net(I_0) \vee \net(g I_0)$, we encounter the same problem that we did in Lemma \ref{Lemma: Almost Covariant to Covariant} stemming from the translation-dilation group being unable to `move points through $\infty$.' With the same argument as in Proposition \ref{Proposition: Cocycle to Endomorphism} it is enough to show that $z_{\dil_I(t)} \in \net(I_0) \vee \net(\dil_I(t) I_0)$ for some $t$ such that $\bar{I}_0 \cap \dil(t) \bar{I}_0 = \emptyset$. 
But this is a direct consequence of the definition of $z_{\dil_I(t)}$. Let a suitable $t$ be given. Then we have for $x \in \net(I \setminus (I_0 \cup \dil_I(t) I_0))$ that
\begin{equation*}
\Ad z_{\dil_I(t)} (x) = \Ad z_{\dil_I(t)} \rho(x) = \alpha_{\dil_I(t)} \circ \rho \circ \alpha_{\dil_I(-t)} (x) = \alpha_{\dil_I(t)} \circ \alpha_{\dil_I(t)} (x) = x .
\end{equation*}
In conclusion, $(z_g)$ is an $\alpha$-cocycle on the translation-dilation group modulo $\rho(\net(I))' \cap \net(I)$ localized in $I_0$. By Remark \ref{Remark: Weak Cocycle to Weakly Translation-Dilation Covariant Endomorphism}, $\rho$ extends to a weakly translation-dilation covariant endomorphism on $S^1$ localized in $I_0$.

Having defined $\rho$ on all of $S^1$, it is easy to see that it is in fact weakly M\"obius covariant. Take a third subinterval $L$ of $S_+$ containing the closure of $I_0$ such that $\dil_L(r)$, $\dil_I(s)$ and $\dil_J(t)$ together generate the M\"obius group. As before we can construct a one-parameter unitary group $t \mapsto U_\rho(\dil_L(t))$ such that $\Ad U_\rho(\dil_L(t)) \circ \rho = \rho \circ \alpha_{\dil_L(t)}$. Checking that $U_\rho(\dil_L(t))$ together with the previously defined representation $U_\rho$ of the group generated by $\dil_I(s)$ and $\dil_J(t)$ generate a representation of the M\"obius group up to left multiplication with $\rho(\net(I))' \cap \net(I)$ is trivial.

Lastly we show that $\sigma^\psi_t \circ \rho = \rho \circ \sigma^\omega_t$. Let $t \in \RR$ be given. For any $t_0$ such that $\dil(t_0) I_0 \bot I$ we have $\rho_{\dil(t_0) I_0^\bot} = \Ad (D \psi : D\omega)_{t_0}$ by the definition of $\rho$. Therefore, if $x \in \net(\dil(t_0) I_0^\bot) \cap \net(\dil(t_0 - t) I_0^\bot) \cap \net(S_+)$, then
\begin{equation*}
\begin{split}
\sigma^\psi_t \circ \rho (x) & = 
\sigma^\psi_t \circ \Ad (D \psi : D\omega)_{t_0 - t} (x)
\\
& = \Ad (D \psi : D\omega)_t \circ \sigma^\omega_t \circ \Ad (D \psi : D\omega)_{t_0 - t} (x)
\\
& = \Ad \prt{ (D \psi : D\omega)_t \sigma^\omega_t ( (D \psi : D \omega)_{t_0 - t})} \sigma^\omega_t(x)
\\
& = \Ad (D \psi: D\omega)_{t_0} \circ \sigma^\omega_t(x)
\\
& = \rho \circ \sigma^\omega_t (x) .
\end{split}
\end{equation*}
This shows that the equation $\sigma^\psi_t \circ \rho = \rho \circ \sigma^\omega_t$ holds on $X_{t_0} := \net(S_+ \setminus ( \dil(t_0) I_0 \cup \dil(t_0 - t) I_0))$. As we can make $X_{t_0}$ cover all of $\net(S_+)$ by varying $t_0$ and $\rho$ does not depend on $t_0$, we conclude that $\sigma^\psi_t \circ \rho = \rho \circ \sigma^\omega_t$ holds on all of $\net(S_+)$.
The equality $\sigma^\psi_t \circ \rho = \rho \circ \sigma^\omega_t$ also ensures uniqueness of $\rho$. 
\end{proof}

Having constructed a weakly M\"obius covariant endomorphism on the basis of a weight, we have completed the first part of the program described in the introduction to this section. For the second part we will use proof elements from \cite{BertozziniContiLongo-CovariantSectors} to show that the endomorphism is a direct sum of finite index endomorphisms. A key element will be played by the following result due to Longo.

\begin{proposition}[\cite{Longo-KacWakimoto}]
\label{Proposition: Direct Sum of Finite Index Factors}
If $\neumannB \subseteq \neumannC$ is an inclusion of factors and if there exists normal faithful conditional expectations $E : \neumannC \rightarrow \neumannB$ and $E' : \neumannC' \rightarrow \neumannB'$, then $\neumannB' \cap \neumannC$ is a direct sum of type I factors. Moreover, for each minimal projection $p$ in $\neumannB' \cap \neumannC$, the inclusion $\neumannB p \subseteq p \neumannC p$ has finite index.
\end{proposition}

\begin{theorem}
\label{Theorem: From Weight to Covariant Endomorphism}
Suppose that $\psi$ is a semi-finite, normal, faithful weight on $\neumannM$ localized in the interval $I_0$, $\bar{I}_0 \subseteq S_+$. There exists a unique consistent endomorphism $\rho$ on $S^1$ localized in $I_0$ such that $\sigma^\psi_t \circ \rho = \rho \circ \sigma^\omega_t$ on $\neumannM$. This $\rho$ is a (possibly infinite) direct sum of finite index endomorphisms. In particular, $\rho$ is M\"obius covariant.
\end{theorem}
\begin{proof}
From Proposition \ref{Proposition: From Weight to Weakly Covariant Endomorphism} we obtain a weakly M\"obius covariant endomorphism $\rho$ localized in $I_0 \subseteq S_+$. Also, as argued in the beginning of the proof of Proposition \ref{Proposition: From Weight to Weakly Covariant Endomorphism}, the Takesaki Theorem provides us with a normal, faithful conditional expectation $E$ of $\net(S_+)$ onto $\rho(\net(S_+))$. We want to construct a normal, faithful conditional expectation from $\net(S_+)'$ onto $\rho(\net(S_+))'$ such that we can utilize Proposition \ref{Proposition: Direct Sum of Finite Index Factors}.

We will write $S_R$ for the right half of the circle $S^1$ corresponding to the interval $]-1;1[$. As the reflection associated with $S_R$ maps $S_+$ bijectively onto $S_+$, the operator $j_{S_R}$ is an anti-automorphism of $\net(S_+)$. Let $\bar{\rho} := j_{S_R} \circ \rho \circ j_{S_R}$. From Proposition \ref{Proposition: Form of Weak Conjugates} we know this to be in the sector of a weak conjugate of $\rho$. As $\bar{\rho}$ is localized in $S_+$, there exists a unitary $u \in \net(S_+)$ such that $\Ad u \circ \bar{\rho} = \rho^{-1} \circ j_{\rho(\neumannM)} j_\neumannM$ on $\net(S_+)$.
Thus, if we define $\bar{E} := j_{S_R} \circ E \circ j_{S_R}$ we get a normal, faithful conditional expectation of $\neumannM$ onto $\bar{\rho}(\neumannM)$. The inclusion $\bar{\rho}(\neumannM) \subseteq \neumannM$ can be written $\Ad (u^*) \circ \rho^{-1} \circ j_{\rho(\neumannM)} j_\neumannM(\neumannM)  \subseteq \neumannM$ which by simple rearrangements is seen to be unitarily equivalent to $\neumannM' \subseteq \rho(\neumannM)'$. In this way we obtain a normal, faithful conditional expectation of $\rho(\neumannM)'$ onto $\neumannM'$ and Proposition \ref{Proposition: Direct Sum of Finite Index Factors} then tells us that $\rho$ is a (possibly infinite) direct sum of finite index endomorphisms. As each of these is automatically M\"obius covariant, so is $\rho$.
\end{proof}

\begin{definition}
Let $\psi$ be a semi-finite, normal, faithful weight on $\neumannM$ localized in some interval $I_0$, $\bar{I}_0 \subseteq S_+$. The unique weakly M\"obius covariant endomorphism $\rho$ on $S^1$ such that $\sigma^\psi_t \circ \rho = \rho \circ \sigma^\omega_t$ is called the \emph{endomorphism associated with $\psi$} and written $\rho_\psi$.
\end{definition}


\subsection{Freedom of Choice of Weights}
The characterizing equation for an endomorphism $\rho$ associated with a weight $\psi$ is $\sigma^\psi_t \circ \rho = \rho \circ \sigma^\omega_t$ or equivalently, 
\begin{equation*}
\Ad \left((D \psi : D\omega)_t\right) \circ \rho = \sigma^\omega_t \circ \rho \circ \sigma^\omega_{-t} .
\end{equation*}
As this only determines the Connes cocycle $(D \psi : D \omega)_t$ up to left multiplication with $\rho(\net(I_0))' \cap \net(I_0)$, assuming $\rho$ localized in $I_0$, many weights will give rise to the same endomorphism. In fact, the following holds.

\begin{proposition}
Let $I_0$ be an interval such that $\bar{I}_0 \subseteq S_+$ and suppose that $\phi$ and $\psi$ are semi-finite, normal, faithful weights on $\neumannM$ localized in $I_0$. Then the following conditions are equivalent:
\begin{enumerate}
\item $(D \psi : D\phi)_t \in \rho_\phi(\neumannM)' \cap \neumannM$, $t \in \RR$.
\item $(D \psi : D\phi)_t \in \rho_\psi(\neumannM)' \cap \neumannM$, $t \in \RR$.
\item $\sigma^\psi_t (x) = \sigma^\phi_t (x)$, $x \in \rho_\phi(\neumannM)$.
\item $\sigma^\psi_t (x) = \sigma^\phi_t (x)$, $x \in \rho_\psi(\neumannM)$.
\item $\rho_\phi = \rho_\psi$.
\end{enumerate}
\end{proposition}
\begin{proof}
As the last condition is symmetric in $\phi$ and $\psi$, it is enough to show that $(1) \Rightarrow (5) \Rightarrow (3) \Rightarrow (1)$.
Assuming $(5)$, then 
\begin{equation*}
\sigma^\psi_t(\rho_\phi(y)) = \sigma^\psi_t( \rho_\psi(y)) = \rho_\psi(\sigma^\omega_t(y)) = \rho_\phi(\sigma^\omega_t(y)) = \sigma^\phi_t (\rho_\phi(y) ).
\end{equation*}
Assuming $(3)$, clearly $(D \psi : D\phi)_t$ commutes with $\rho_\phi(\neumannM)$ as stated in $(1)$.
Finally, we assume $(1)$. As both $\rho_\phi$ and $\rho_\psi$ are localized in $I_0$, it is enough to show that they coincide on $\net(I_0)$. Choose $t_0$ such that $\dil(t_0) I_0 \bot I_0$. Then
\begin{equation*}
\begin{split}
\rho_\phi|_{\net(I_0)} &= \Ad (D \psi : D\phi)_{t_0} \circ \rho_\phi|_{\net(I_0)}
\\
& = \Ad (D \psi : D\phi)_{t_0} \circ \Ad (D \phi : D\omega)_{t_0}|_{\net(I_0)}
\\
& = \Ad (D \psi : D\omega)_{t_0}|_{\net(I_0)} 
\\
& = \rho_\psi|_{\net(I_0)}
\end{split}
\end{equation*}
as per condition $(5)$.
\end{proof}

In case the endomorphism $\rho_\psi$ is irreducible, the weight $\psi$ is unique up to a scalar:

\begin{corollary}
Let $\rho$ be an irreducible M\"obius covariant endomorphism localized in $I_0$, $\bar{I}_0 \subseteq S_+$. If $\psi$ and $\rho$ are weights localized in $I_0$ such that $\rho_\psi = \rho_\phi = \rho$ then $\psi = \lambda \phi$ for some positive real $\lambda$.
\end{corollary}
\begin{proof}
As $(D \psi : D \phi)_t \in \rho(\neumannM)' \cap \neumannM = \CC$, there exists a positive scalar $\lambda$ such that $\lambda^{it} = (D \psi : D \phi)_t$ for all $t \in \RR$. By the defining properties of the Connes cocycle derivative, $\psi = \lambda \phi$.
\end{proof}

While in general there is a certain amount of freedom in the choice of the weight inducing a given endomorphism $\rho$, we want to single out the weight $\psi$ satisfying $(D \psi : D\omega)_t = z_\rho(\dil(t))$. Bertozzini, Conti and Longo have given a criterion for when this is satisfied for finite index endomorphisms:

\begin{proposition}[\cite{Longo-KacWakimoto,BertozziniContiLongo-CovariantSectors}]
\label{Proposition: Good Form of Weights}
Let $\rho$ be a M\"obius covariant endomorphism localized in $I_0$, $\bar{I}_0 \subseteq S_+$ and $\psi$ a positive linear functional on $\neumannM$. Then the following conditions are equivalent.
\begin{enumerate}
\item $\psi$ is normal and faithful and its Connes cocycle derivative relative to $\omega$ satisfies $(D \psi : D\omega)_t = z_\rho(\dil(t))$, $t \in \RR$.
\item $\psi = \ind(\rho) \omega \circ \rho^{-1} \circ E_\rho$ where $E_\rho$ is the minimal conditional expectation of $\neumannM$ onto $\rho(\neumannM)$.
\item $\psi(xy^*) = \ip{\exp(-D_\rho /2) x \Omega}{\exp(-D_\rho /2) y \Omega}$ where $D_\rho$ is the infinitesimal generator of $t \mapsto U_\rho(\dil(t))$.
\item $\psi$ is normal, faithful, satisfies $\sigma^\psi_t \circ \rho = \rho \circ \sigma^\omega_t$, $t \in \RR$, and $\psi|_{\rho(\neumannM)' \cap \neumannM}$ is a trace whose value on a central projection $p$ is given by $\psi (p) = \ind(\rho_p)$ where $\rho_p$ is the subrepresentation of $\rho$ associated with $p$.
\end{enumerate}
\end{proposition}

The condition that $\psi$ is a trace on $\rho(\neumannM)' \cap \neumannM$ is equivalent to $\rho(\neumannM)' \cap \neumannM$ being a subset of the centralizer of $\psi$. This can be restated as the condition that the modular automorphism group of the conditional expectation $E_\rho$ is trivial.

For later use we also mention that condition (2) can be rewritten as $\psi \circ \rho = \ind(\rho) \omega$, suggesting that it may be useful to think of $\psi$ as the vacuum state for the net $(\rho(\net(I)))$.


\section{Basic Constructions with Weights}
\label{Section: Basic Constructions with Weights}

In order to work with weights instead of endomorphisms, we need to establish the analogues of unitary equivalence, direct sums, conjugate endomorphisms and the like within the framework of weights. In this section we take the first basic steps in that direction. 

As usual we fix a standard half-sided modular inclusion $(\neumannN \subseteq \neumannM, \Omega)$ and denote the associated net of von Neumann algebras by $\net$.

\subsection{Unitary Equivalence, Subrepresentations and Direct Sums}
We introduce the notions of unitary equivalence, subrepresentations and direct sums for weights. Similar considerations appear in \cite{Longo-KacWakimoto}.

\begin{remark}[The Connes Cocycle Derivative]
\label{Remark: The Connes Cocycle Derivative}
We briefly recall some facts and notation concerning the Connes cocycle derivative. Define $\band := \{z \in \CC \mid 0 \leq \Imag z \leq 1 \}$ and let $\anal(\band)$ denote the set of bounded, continuous functions on $\band$ that are analytic in the interior. If $\psi$ is a weight on $\neumannM$ we define $\n_\psi := \{ x \in \neumannM \mid \psi(x^* x) < \infty \}$. 

For given normal, semi-finite, faithful weights $\psi$ and $\phi$ on $\neumannM$, the Connes cocycle derivative $(D \psi : D\phi)_t$ is uniquely determined by the following condition \cite{Takesaki2}: For each $x \in \n_\psi \cap \n_\phi^*$ and $y \in \n_\psi^* \cap \n_\phi$, there exists a function $F_{x,y} \in \anal(\band)$ such that
\begin{equation*}
F(t) = \psi \prt{ (D \psi : D \phi)_t \sigma^\phi_t(y) x} , \quad F(t+i) = \phi \prt{ x (D \psi: D \phi)_t \sigma^\phi_t(y) }; \quad t \in \RR .
\end{equation*}
\end{remark}

\begin{proposition}
\label{Proposition: Unitary Equivalence for Weights}
Suppose that $\psi$ is a semi-finite, normal, faithful weight localized in the interval $I_0$, $\bar{I}_0 \subseteq S_+$ and that $u$ is a unitary localized in some interval whose closure is contained in $S_+$. Then $\rho_\psi$ is unitarily equivalent to $\rho_{\psi_u}$ where $\psi_u := \psi \circ \Ad u^*$.
\end{proposition}
\begin{proof}
We will write $\rho$ for the endomorphism $\rho_\psi$ and $\rho_u$ for the endomorphism $\Ad u \circ \rho$. Defining $U_{\rho_u}(g) = u U_\rho(g) u^*$ gives a representation of the universal covering group of the M\"obius group satisfying $\Ad U_{\rho_u}(g) \circ \rho_u = \rho_u \circ \alpha_g$. We define $\psi_u$ to be the weight whose Connes cocycle derivative with respect to $\omega$ satisfies $(D \psi_u : D\omega)_t = U_{\rho_u}(\dil(t)) U(\dil(t))^*$. Thus, $(D \psi_u : D \omega)_t = u (D \psi : D\omega)_t \sigma^\omega_t(u)^*$. 

For $x \in \n_{\psi_u} \cap \n_\omega^*$ and $y \in \n_\omega \cap \n_{\psi_u}^*$ we have an $F_{x,y} \in \anal(\band)$ such that
\begin{equation*}
F_{x,y}(t) = \psi_u(u (D \psi: D\omega)_t \sigma^\omega_t(u^* y) x) , \quad F_{x,y}(t+i) = \omega(x u (D \psi: D\omega)_t \sigma^\omega_t(u^*y) ) .
\end{equation*}
We recognize $F_{x,y}(t+i)$ as being equal to the function $G_{xu, u^*y}(t+i)$ where $G \in \anal(\band)$ is the unique function determining $(D \psi: D\omega)_t$, i.e., $G_{xu, u^*y}(t) = \omega((D \psi: D\omega)_t \sigma^\omega_t(u^*y) xu)$. In particular,
\begin{equation*}
\psi_u(yx) = F_{x,y}(0) = G_{xu,u^*y}(0) = \psi(u^* y x u)
\end{equation*}
and we conclude that $\psi_u = \psi \circ \Ad u^*$.
\end{proof}

The proof above would work just as well, were the unitary $u$ to be replaced by an isometry. This observation leads to the following two propositions.

\begin{proposition}
Suppose that $\psi$ is a semi-finite, normal, faithful weight localized in the interval $I_0$, $\bar{I}_0 \subseteq S_+$. If $\rho$ is a subrepresentation of $\rho_\psi$, $\rho = \Ad v \circ \rho_\psi$, then $\rho = \rho_{\psi \circ \Ad v^*}$.
\end{proposition}
\begin{proof}
The proof is identical to that of Proposition \ref{Proposition: Unitary Equivalence for Weights}.
\end{proof}

\begin{proposition}
Suppose that $\psi_1, \ldots, \psi_n$ are semi-finite, normal, faithful weights localized in $I_0$. For any choice of isometries $v_1, \ldots, v_n \in \net(I_0)$ such that $\sum v_i v_i^* = 1$, the weight $\psi := \sum \psi_i \circ \Ad v_i^*$ satisfies $\rho_\psi = \sum \Ad v_i \circ \rho_{\psi_i}$.
\end{proposition}
\begin{proof}
The proof is a slight variation of that of Proposition \ref{Proposition: Unitary Equivalence for Weights}.
\end{proof}

In case we are considering weights of the form described in Proposition \ref{Proposition: Good Form of Weights}, i.e., weights $\psi$ satisfying $(D \psi : D\omega)_t = z_{\rho_\psi}(\dil(t))$, it is easy to check that forming unitary equivalence, subrepresentations and direct sums preserves this form.


\subsection{Weak Conjugates for Weights}

While several representatives of a weak conjugate of a given endomorphism $\rho$ localized in $S_+$ exists, we focus on the most obvious one which is likewise localized in $S_+$, namely $j_{S_+} \circ \rho \circ j_{S_+}$. Longo presents similar considerations in \cite{Longo-KacWakimoto}.

\begin{proposition}
Suppose that $\psi$ is a semi-finite, normal, faithful weight localized in the interval $I_0$, $\bar{I}_0 \subseteq S_+$, and that $r$ is the reflection associated with the right half-circle $S_R$. Then the endomorphism associated with the weight $\bar{\psi} := \psi \circ j_r$ localized in $r I_0 \subseteq S_+$ is a weak conjugate of $\rho_\psi$, in fact $\rho_{\bar \psi} = j_r \circ \psi \circ j_r$.
\end{proposition}
\begin{proof}
By Proposition \ref{Proposition: Form of Weak Conjugates}, the endomorphism $j_r \circ \rho_\psi \circ j_r$ is a weak conjugate of $\rho_\psi$, and clearly $j_r \circ \rho_\psi \circ j_r$ is localized in $r I_0 \subseteq S_+$. The $\alpha$-cocycle associated with $j_r \circ \rho_\psi \circ j_r$ is $\bar{z}_g = j_r (z_{rgr})$. We seek the weight $\bar{\psi}$ whose Connes cocycle derivative with respect to $\omega$ is $\bar{z}_{\dil(t)}$. As usual we do this by computing the function $F^{\bar{\psi}} \in \anal(\band)$ satisfying the following for $x \in \n_{\bar{\psi}} \cap \n_\omega^*$ and $y \in \n_\omega \cap \n_{\bar{\psi}}^*$,
\begin{equation*}
F^{\bar{\psi}}_{x,y} (t) = \bar{\psi}(\bar{z}(\dil(t)) \sigma^\omega_t(y) x) ,
\quad F^{\bar{\psi}}_{x,y}(t+i) = \omega(x \bar{z}(\dil(t)) \sigma^\omega_t(y)) .
\end{equation*}
It is easy to calculate that $r \dil(t) r = \dil(-t)$ and that consequently $j_r \circ \sigma^\omega_t \circ j_r = \sigma^\omega_{-t}$. Using this and the fact that $\omega = \omega \circ j_r$, we can rewrite $F^{\bar{\psi}}_{x,y}(t+i)$ as
\begin{equation*}
\begin{split}
F^{\bar{\psi}}_{x,y}(t+i) & = \omega( x j_r(z_{\dil(-t)}) \sigma^\omega_t(y) )
\\
& = \omega (j_r( x j_r(z_{\dil(-t)}) j_r(\sigma^\omega_{-t}(j_r(y))))) 
\\
& = \omega ( j_r(x) z_{\dil(-t)} \sigma^\omega_{-t} (j_r(y)) )
\\
& = F^\psi_{j_r(x), j_r(y)}(-t + i) .
\end{split}
\end{equation*}
Thus we find that 
\begin{equation*}
\bar{\psi}(yx) = F^{\bar{\psi}}_{x,y}(0) = F^\psi_{j_r(x),j_r(y)}(0) = \psi(j_r(yx)) 
\end{equation*}
and conclude that $\bar \psi = \psi \circ j_r$.
\end{proof}

We note in passing that if the weight $\psi$ is of the form given in Proposition \ref{Proposition: Good Form of Weights}, then $\bar \psi$ is likewise of that form.


\subsection{Criteria for Finite Index}

It is possible for a given weight $\psi$ to determine whether or not $\rho_\psi$ has finite index on the basis of the weight alone. In fact, Longo has shown that if the weight $\psi$ is of the form given in Proposition \ref{Proposition: Good Form of Weights}, then the index of $\rho_\psi$ is $\ind(\rho_\psi) = \psi(1)$ \cite{Longo-KacWakimoto}. In particular, $\rho_\psi$ has finite index if and only if $\psi$ is a functional.

For a general weight $\psi$ localized in an interval $I_0$ it is more difficult to determine the index of $\rho_\psi$ directly. It is easy to construct a weight $\psi$ such that $\rho_\psi$ has infinite index but $\psi(1) < \infty$. 

\begin{proposition}
Let $\psi$ be a normal, semi-finite, faithful weight localized in $I_0$, $\bar{I}_0 \subseteq S_+$. The following holds.
\begin{enumerate}
\item If any normal, semi-finite, faithful weight $\phi$ localized in $I_0$ for which $\sigma^\phi|_{\rho_\psi(\neumannM)} =\sigma^\psi|_{\rho_\psi(\neumannM)}$  and $\rho(\neumannM)' \cap \neumannM \subseteq \neumannM_\phi$ is a functional, then $\rho_\psi$ has finite index.
\item If $\psi$ is a functional and $\psi \circ \rho_\psi$ is proportional to $\omega$, then $\rho_\psi$ has finite index.
\end{enumerate}
\end{proposition}
\begin{proof}
If $\psi_0$ is the weight for which the Connes cocycle derivative satisfies $(D \psi_0 : D \omega)_t = U_{\rho_\psi}(\dil(t))$ then $\rho_\psi$ has finite index if and only if $\phi_0$ is a functional \cite{Longo-KacWakimoto}. Condition (2) is simply a restatement of $\psi_0$ being a functional, see the comment after Proposition \ref{Proposition: Good Form of Weights}.

As concerns the first condition, if $\phi$ is a normal, semi-finite, faithful weight localized in $I_0$ such that $\sigma^\phi|_{\rho_\psi(\neumannM)} = \sigma^\psi|_{\rho_\psi(\neumannM)}$ and $\rho_\psi(\neumannM)' \cap \neumannM \subseteq M_\phi$, then $\rho_\phi = \rho_\psi$ and $\sigma^\phi = \sigma^{\psi_0}$. The latter equality is equivalent to $\phi = \lambda \psi_0$ for some real number $\lambda >0$. Hence, $\phi$ is a functional if and only $\psi_0$ is a functional, and $\psi_0$ is a functional if and only if $\rho_\psi = \rho_{\psi_0}$ has finite index \cite{Longo-KacWakimoto}.
\end{proof}

\subsection{Positive Energy}
As any endomorphism $\rho_\psi$ associated with a weight as per Theorem \ref{Theorem: From Weight to Covariant Endomorphism} is a direct sum of finite index endomorphisms, and any endomorphism of finite index has positive energy we can make the following easy observations.

\begin{proposition}
Suppose that $\psi$ is a normal, semi-finite, faithful weight localized in $I_0$, $\bar{I}_0 \subseteq S_+$. Then the associated endomorphism $\rho_\psi$ has positive energy.
\end{proposition}

\begin{corollary}
Suppose that $\rho$ is a M\"obius covariant endomorphism localized in $I_0$, $\bar{I}_0 \subseteq S_+$ and that $\psi$ is the weight on $\neumannM$ for which $(D \psi : D \omega)_t = z_\rho(\dil(t))$. If it holds that $\psi |_{\rho(\neumannM)}$ is semi-finite, then $\rho$ has positive energy.
\end{corollary}

Other criteria for a M\"obius covariant endomorphism having positive energy can be found in \cite{BertozziniContiLongo-CovariantSectors}.



\end{document}